\documentclass[fleqn,usenatbib]{mnras}

\usepackage{newtxtext,newtxmath}
\usepackage{multirow}

\usepackage[T1]{fontenc}
\usepackage{amsmath} % or simply amstext

\DeclareRobustCommand{\VAN}[3]{#2}
\let\VANthebibliography\thebibliography
\def\thebibliography{\DeclareRobustCommand{\VAN}[3]{##3}\VANthebibliography}

\usepackage{graphicx}	% Including figure files
\usepackage{amsmath}	% Advanced maths commands
\title[HR 6819: A stripped star + Be star binary]{A stripped-companion origin for Be stars: clues from the putative black holes HR 6819 and LB-1}

\author[El-Badry \& Quataert]{
Kareem El-Badry$^{1}$\thanks{E-mail: kelbadry@berkeley.edu}
and Eliot Quataert$^{1}$ \\
$^{1}$Department of Astronomy and Theoretical Astrophysics Center, University of California Berkeley, Berkeley, CA 94720, USA\\
}

\date{Accepted XXX. Received YYY; in original form ZZZ}

\pubyear{2020}

\begin{document}
\label{firstpage}
\pagerange{\pageref{firstpage}--\pageref{lastpage}}
\maketitle

\begin{abstract}
HR 6819 is a bright ($V=5.36$), blue star recently proposed to be a triple containing a detached black hole (BH). We show that the system is a binary and does not contain a BH. 
Using spectral decomposition, we disentangle the observed composite spectra into two components: a rapidly rotating Be star and a slowly rotating B star with low surface gravity $(\log g \approx 2.75)$. Both stars show periodic radial velocity (RV) variability, but the RV semi-amplitude of the B star's orbit is $K_{\rm B}= (62.7 \pm 1)\,\rm km\,s^{-1}$, while that of the Be star is only $K_{\rm Be} = (4.5\pm 2)\,\rm km\,s^{-1}$. This implies that the B star is less massive by at least a factor of 10.  The surface abundances of the B star bear imprints of CNO burning. We argue that the B star is a bloated, recently stripped helium star with mass $\approx 0.5\,M_{\odot}$ that is currently contracting to become a hot subdwarf.
The orbital motion of the Be star obviates the need for a BH to explain the B star's motion. 
A stripped-star model reproduces the observed luminosity of the system, while a normal star with the B star's temperature and gravity would be more than 10 times too luminous. 
HR 6819 and the binary LB-1 probably formed through similar channels. We use MESA models to investigate their evolutionary history, finding that they likely formed from intermediate-mass ($3-7\,M_{\odot}$) primaries stripped by slightly lower-mass secondaries and are progenitors to Be + sdOB binaries such as $\phi$\,Persei. The lifetime of their current evolutionary phase is on average $2\times 10^5$\,years, of order half a percent of the total lifetime of the Be phase. This implies that many Be stars have hot subdwarf and white dwarf companions, and that a substantial fraction ($20-100\%$) of field Be stars form through accretion of material from a binary companion. 
\end{abstract}


%%%%%%%%%%%%%%%%%%%%%%%%%%%%%%%%%%%%%%%%%%%%%%%%%%

%%%%%%%%%%%%%%%%% BODY OF PAPER %%%%%%%%%%%%%%%%%%

\section{Introduction}

%Interactions between the members of binary and multiple star systems give rise to a wealth of imperfectly understood physical phenomena and produce many of the most exotic astrophysical objects. Because the gravitational effects of a massive object often produce observable changes in its companion, binary and multiple star systems are also prime targets for searches for faint objects such as neutron stars and black holes (BHs). 
Large-scale multi-epoch radial velocity (RV) surveys have identified a number of binaries in recent years that are proposed to contain stellar mass black holes \citep[BHs; e.g.][]{Casares_2014, Khokhlov_2018, Giesers_2018, Giesers_2019, Thompson_2019}. Confirmation of these BH candidates is challenging precisely because BHs are expected to be rare. That is, all plausible alternate explanations for observed data, even those which are rare, must be ruled out before a candidate BH can be considered reliable.

Recently, \citet{Rivinius_2020} identified the object HR 6819 as a candidate host of a stellar-mass BH. Phase-resolved optical spectra of the object revealed two luminous components: a B star with relatively narrow, RV-variable absorption lines, which were observed to follow a nearly circular orbit with $P= 40.3$ days and velocity semi-amplitude $K_{\rm B}\approx 61\,\rm km\,s^{-1}$, and a classical Be star whose broad emission and absorption lines appeared to be stationary. No component was found to orbit in anti-phase with the B star. Assuming the mass of the B star to be at least $5\,M_{\odot}$, as expected for a normal star of its spectral type, \citet{Rivinius_2020} argued that any stellar companion massive enough to explain the B star's orbit would also contribute to the spectrum at a detectable level. They thus concluded that the companion is a BH, with an estimated minimum mass of $4.2\,M_{\odot}$. In this hierarchical triple scenario, the Be star must be at least a few AU from the B star and BH for the system to be dynamically stable, and its status as a Be star would likely be unrelated to the B star or BH.

HR 6819 is in many ways similar to the binary LB-1 \citep{Liu_2019}, which was also proposed to contain a stellar-mass BH. Like HR 6819, LB-1 contains an RV-variable B star and apparently stationary emission lines. The emission lines were initially proposed to originate in an accretion disk, either around the BH \citep{Liu_2019} or around the binary \citep{ElBadry_2020, AbdulMasih_2020, Irrgang_2020}.
\citet{Rivinius_2020} proposed that LB-1 and HR 6819 are both hierarchical triples with a B star and a BH companion in the inner binary, and a distant Be star -- the source of the emission lines -- orbiting both components. 

\citet{Shenar_2020}, however, recently used spectral disentangling to fit the multi-epoch spectra of LB-1 as a sum of two luminous components. They also found evidence for a Be star, including both emission and rotationally-broadened absorption lines, but found its spectrum to shift in anti-phase with the B star. This is not expected in the hierarchical triple scenario, so they argued that LB-1 is a binary containing two luminous stars. It was also recently noted by \citet{Liu_2020} that the emission line shape in LB-1 varies coherently with the B star's phase. This variation, which is consistent with expectations for an irradiated or tidally-perturbed disk, is expected if the system is a binary, but not if it is a hierarchical triple. 

The RV variability amplitude found by \citet{Shenar_2020} for the Be star in LB-1 suggests it is 5 times more massive than the B star. If the luminous binary scenario for LB-1 is correct, the B star thus would have a mass of $ M_{\rm B} \approx 1.5\,M_{\odot}$, a factor of 4 lower than expected for a normal star of its spectral type. \citet{Shenar_2020} propose that most of the B star's envelope was recently stripped by its companion. In this case, the star is currently contracting and will likely soon become a core helium burning sdOB star (see also \citealt{Irrgang_2020}). 

In this paper, we use spectral disentangling to fit the multi-epoch spectra of HR 6819. We find that, like in LB-1, the Be star orbits in anti-phase with the B star, which is much less massive than expected for a normal star of its spectral type. The remainder of this paper is organized as follows. We describe the spectra in Section~\ref{sec:data} and the disentangling method in Section~\ref{sec:disentangle}. We constrain the atmospheric parameters of both components in Section~\ref{sec:teff}. Section~\ref{sec:velocites} presents the RV variability of both components, and Section~\ref{sec:lum} compares the system's luminosity to models. The emission from the Be star's disk is examined in Section~\ref{sec:emission}, and the abundances of the B star in Section~\ref{sec:abundances}. We present possible evolutionary models for the system in Section~\ref{sec:evol}. Finally, we discuss HR 6819 and LB-1 in the context of the broader Be star population in Section~\ref{sec:discussion}. 

The appendices provide supporting information. 
Appendix~\ref{sec:distance_anchor} describes our estimate of both components' masses, radii, and luminosities. Spectral disentangling is detailed in Appendix~\ref{sec:disentangle_details}. Appendix~\ref{sec:variability} investigates the pulsation-driven variability of the B star. Spectroscopic constraints on the B star's rotation velocity are presented in Appendix~\ref{sec:rot_vs_vmac}.  Its surface helium abundance is investigated in Appendix~\ref{sec:He_enrichment}.

\section{Methods}

\subsection{Data}
\label{sec:data} % used for referring to this section from elsewhere

We analyze 51 optical spectra of HR 6819 with spectral resolution $R\approx 48,000$ that were taken in 2004 with the FEROS echelle spectrograph \citep{Kaufer_1999} on the ESO/MPG 2.2m telescope at La Silla Observatory. The data span 134 days (3.3 orbital periods) from MDJ 53138 to 53273 and are publicly available through the ESO archive; they are described in more detail by \citet{Rivinius_2020}. The spectra were reduced with the ESO-MIDAS pipeline, which performs bias-subtraction, flat fielding, and wavelength calibration, applies a heliocentric correction, and combines the spectra from individual orders. The pipeline failed to reliably measure the signal-to-noise ratio (SNR), so we estimated it empirically from the pixel-to-pixel scatter in regions without strong absorption or emission lines. The typical SNR per pixel is 300 at $\lambda = 6,000\,\textup{\AA}$.\footnote{All wavelengths are quoted at their values in air.} We verified the stability of the wavelength solution by checking that the ISM sodium absorption line at 5890 \textup{\AA} is found at the same heliocenter-corrected wavelength at all epochs. 

We also analyzed an additional 12 FEROS spectra of HR 6819 that were obtained in 1999 (dataset ``A'' in \citealt{Rivinius_2020}). We found that the Be star's emission line profiles varied substantially between the 1999 and 2004 datasets, with obvious changes in the shape, width, and amplitude of Balmer, Fe II, and O I emission lines. Such variation is common in Be stars due to changes in the structure of the disk \citep[e.g.][]{Dachs_1981, Okazaki_1991}, but it complicates the disentangling of the composite spectra. We therefore focus our analysis on the 2004 dataset, within which the emission line profiles are relatively stable.

The spectra are not flux calibrated. Continuum normalization was performed by fitting a cubic spline to wavelength pixels without significant absorption or emission. The initial set of candidate continuum pixel was selected by taking all wavelengths from a TLUSTY model spectrum \citep{Hubeny_1995} with $T_{\rm eff} = 18\rm \, kK$ and $\log g=3.5$ for which all pixels within $\pm 60\,\rm km\,s^{-1}$ are within 0.5\% of the theoretical continuum. We then refined the selection of continuum pixels by inspecting the individual spectra, removing pixels affected by telluric absorption and other features not present in the model spectrum. We tested application of this continuum normalization procedure on mock spectra with a range of $\log g$ to verify that it does not significantly overfit the continuum in the wings of broad lines. This is important, because improper continuum normalization can change the shape of the wings of broad lines, leading to biases in the inferred  $\log g$. We focus our analysis on the part of the spectrum with $\lambda > 3900\,\textup{\AA}$ because the narrow spacing of the Balmer lines at shorter wavelengths makes it challenging to estimate the continuum there. 

\subsection{Spectral disentangling}
\label{sec:disentangle}
We used the code \texttt{CRES} \citep{Ilijic_2004}, which implements the wavelength-space spectral disentangling algorithm of \citet{Simon_1994}, to separate the spectra of the two components. The algorithm operates under the ansatz that all the observed composite spectra are produced by summing together the time-invariant spectra of two components, with known Doppler shifts between epochs and a known continuum luminosity ratio. The rest-frame spectra of the two components are modeled as arbitrary vectors, which under these assumptions can be solved for using singular value decomposition. No model spectra (or indeed, any physics beyond the Doppler shift) are use in this calculation. 

The velocity of the B star in HR 6819 is reasonably well known at each epoch through the orbital solution from \citet{Rivinius_2020}, but the velocity of the Be star is not. Following \citet{Shenar_2020}, we therefore proceed under the ansatz that the Be star and B star are orbiting each other, and step through a grid of velocity semi-amplitudes for the Be star, $K_{\rm Be}$, calculating the best-fit disentangled spectra for each value. The choice of $K_{\rm Be}=0\,\rm km\,s^{-1}$ corresponds to a stationary Be star and would be expected in a triple scenario. We choose the optimal $K_{\rm Be}$ and spectral decomposition as the value that minimizes the total $\chi^2$ over all spectra; we find $K_{\rm Be} = (4.5\pm 2)\,\rm km\,s^{-1}$. The determination of this value is described in detail in Appendix~\ref{sec:disentangle_details}. 

When determining $K_{\rm Be}$, we use the spectra at their native resolution and independently analyze narrow wavelength ranges centered on absorption and emission lines with clear contributions from both components. We then adopt $K_{\rm Be}=4.5\,\rm km\,s^{-1}$ and use this value to disentangle the full spectrum. At this stage, we re-bin the spectra to a $0.2\,\rm \textup{\AA}$ wavelength resolution to reduce the computational cost of disentangling. This does not significantly decrease the information content of the spectra, because higher-frequency spectral features are smeared out by rotation and/or macroturbulence.  The spectral disentangling and determination of $K_{\rm Be}$ are described in detail in Appendix~\ref{sec:disentangle_details}.

Spectral decomposition only determines the spectra of the two components up to a multiplicative constant, which represents the continuum luminosity ratio of the two stars. Unless there are  e.g. eclipses or well-characterized pulsations, the luminosity ratio must be determined based on external information, because the effects on the combined spectrum of increasing the continuum flux contribution of one component are identical to the effects of weakening all its spectral lines. We begin by assuming both components contribute 50\% of the light, as estimated by \citet{Rivinius_2020}. Once a preliminary  decomposition is found with this ratio, we adjust the luminosity ratio to match the depths of the B star's Balmer lines to model spectra, while accounting for rotation, macroturbulence, instrumental broadening, and rebinning. The temperature of the model spectra is constrained primarily by ionization balance of metal lines (see Section~\ref{sec:teff}; this temperature estimate is not sensitive to the luminosity ratio).   We found the optimal flux ratio to be $f_{\rm B}/f_{\rm tot}=0.47\pm 0.05$ at $\lambda = 4000\,\textup{\AA}$ (i.e., with the Be star the slightly brighter component). The uncertainty in the flux ratio accounts for uncertainty in $T_{\rm eff}$ and $\log g$, as all atmospheric parameters are fit simultaneously. Because the Be star is slightly hotter, its relative contribution to the total flux is expected to decrease toward redder wavelengths. For our fiducial temperature estimates of $T_{\rm eff, B}=16\,\rm kK$ and $T_{\rm eff, Be}=18\,\rm kK$ (Section~\ref{sec:teff}), the expected flux ratio increases to 0.49/0.51 at $\lambda = 7000\,\textup{\AA}$, still with the Be star the brighter component. This weak wavelength-dependence of the flux ratio is accounted for in our spectral disentangling. We normalize the component spectra following the method described by \citet{Ilijic_2004b}.

\begin{figure*}
	\includegraphics[width=\textwidth]{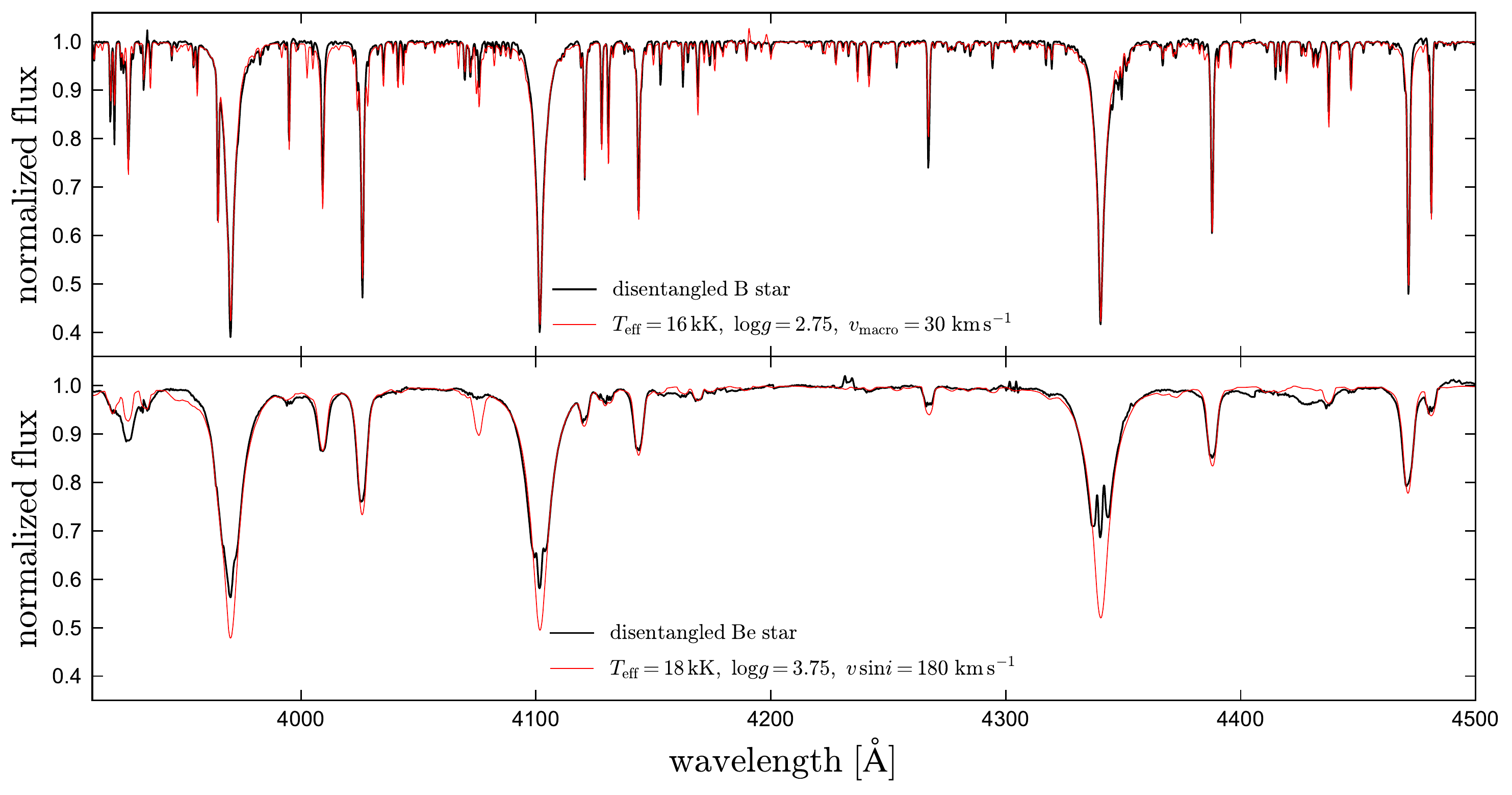}
    \caption{Disentangled spectra of the B star (top) and Be star (bottom) in HR 6819. Black lines show the reconstructed spectra. Red lines show TLUSTY model spectra, which are not used in the spectral disentangling, for comparison. The B star spectrum contains many narrow lines and is well-fit by a model with $T_{\rm eff}=16\,\rm kK$ and $\log g = 2.75$ (see also Figure~\ref{fig:balmer}). The Be star spectrum is strongly broadened by rotation. It is reasonably well described by a model with $T_{\rm eff}=18\,\rm kK$ and $\log g = 3.75$, but with clear excess emission inside the Balmer line cores (see also Figure~\ref{fig:emission_lines}).}
    \label{fig:disentangled_spectra}
\end{figure*}

Figure~\ref{fig:disentangled_spectra} shows a $600\,\textup{\AA}$ wide section of the disentangled spectra. The overplotted TLUSTY model spectra are taken from the BSTAR06 grid \citep{Lanz_2007} and have the $T_{\rm eff}$, $\log g$, and $v\,\sin i$ that we find to best fit the disentangled spectra. The chemical abundances of the models are derived in Section~\ref{sec:abundances}. The Be star spectrum primarily shows strong absorption lines from neutral H and He due to its large $v\,\sin i$, as well as excess due to double-peaked emission lines within the Balmer line cores and in Fe II lines (Section~\ref{sec:emission}). The B star spectrum has many narrow metal lines (see Section~\ref{sec:abundances}).

Fitting the disentangled metal line profiles of both components with the rotational profile from \citet{Gray_1992}, we find $v\sin i = (180\pm 20)\,\rm km\,s^{-1}$ for the Be star. For our best-fit Be star mass, radius, and inclination (Sections~\ref{sec:velocites} and~\ref{sec:lum}), this corresponds to near-critical rotation. For the B star, we derive an upper limit of $v\sin i < 20\,\rm km\,s^{-1}$. We are not able to measure the actual value of $v \sin i$ from the B star's line profiles because non-rotational broadening due to macroturbulence and/or non-radial pulsations dominates over rotation (see Appendix~\ref{sec:rot_vs_vmac}). If the B star is tidally synchronized, as might be expected following a period of mass transfer, its $v \sin i$ would be less than $5\,\rm km\,s^{-1}$. The Be star is not expected to be tidally synchronized, as accretion of mass and angular momentum is expected to spin it up. 

Overall, the consistency between the disentangled spectra and theoretical models, as manifest in properties such as the relative depth and width of absorption lines, is good. We emphasize that the spectral disentangling procedure does not use model spectra in any way -- the disentangled spectra are free to take any shape they like -- so the agreement between the observed and model spectra is encouraging.

\subsection{Temperature and gravity}
\label{sec:teff}

\begin{table}
\centering
\caption{Physical parameters and 1$\sigma$ (middle 68\%) uncertainties for both components of HR 6819. Constraints on stellar parameters are based on the measured $T_{\rm eff}$ and $\log g$ of the Be star and the dynamical mass ratio. Table~\ref{tab:distance_and_tracks} lists constraints that also take the distance into account. }
\begin{tabular}{lll}
\hline\hline
\multicolumn{3}{l}{\bf{Parameters of the B star}}  \\ 
Effective temperature & $T_{\rm eff}$\,[kK] & $16 \pm 1$ \\
Surface gravity   & $\log(g/(\rm cm\,s^{-2}))$  & $2.75\pm0.35$  \\
Projected rotation velocity & $v\sin i$\,[km\,s$^{-1}$] &  $< 20$ \\
Macroturbulent velocity & $v_{\rm macro}$\,[km\,s$^{-1}$] & $30\pm 5$ \\
Microturbulent velocity & $v_{\rm mic}$\,[km\,s$^{-1}$] & $10\pm 5$ \\

Continuum flux ratio at 4000\,\textup{\AA} & $f_{{\rm B}}/f_{{\rm tot}}(4000\,\text{\AA})$ & $0.47\pm0.05$ \\
Radius & $R\,[R_{\odot}]$ & $4.7^{+2.9}_{-1.9}$  \\ 
Bolometric luminosity & $\log(L/L_{\odot})$ & $3.11^{+0.42}_{-0.46}$ \\ Mass &  $M\,[M_{\odot}]$ & $0.47^{+0.28}_{-0.22}$ \\

\hline
\multicolumn{3}{l}{\bf{Parameters of the Be star}}   \\ 
Effective temperature & $T_{\rm eff}$\,[kK] & $18^{+2}_{-3}$ \\
Surface gravity   & $\log(g/(\rm cm\,s^{-2}))$  & $3.75^{+0.5}_{-0.25}$  \\
Projected rotation velocity & $v\sin i$\,[km\,s$^{-1}$] &  180 $\pm$ 20 \\
Macroturbulent velocity & $v_{\rm macro}$\,[km\,s$^{-1}$] & $60\pm 20$ \\
Continuum flux ratio at 4000\,\textup{\AA} & $f_{{\rm Be}}/f_{{\rm tot}}(4000\,\text{\AA})$ & $0.53\pm0.05$ \\
Mass &  $M\,[M_{\odot}]$ & $6.7^{+1.9}_{-1.5}$ \\ 
Radius & $R\,[R_{\odot}]$ & $4.7^{+2.7}_{-1.5}$  \\ 
Bolometric luminosity & $\log(L/L_{\odot})$ & $3.35^{+0.47}_{-0.44}$ \\ 
Fraction of critical rotation & $v_{\rm rot}/v_{\rm crit}$ & $0.80^{+0.18}_{-0.14}$ \\ 

\hline
\multicolumn{3}{l}{\bf{Parameters of the binary}}   \\ 
Orbital period & $P$\,[day]  & $40.3\pm0.3$  \\
B star RV semi-amplitude  & $K_{\rm B}$ [km\,s$^{-1}$] & 62.7$\pm$1.0 \\
Be star RV semi-amplitude  & $K_{\rm Be}$ [km\,s$^{-1}$] & 4.5$\pm$2.0 \\
Be star center-of-mass velocity & $\gamma_{\rm B}$\,[km\,s$^{-1}$] & $10.4\pm 0.6$ \\ 
Be star center-of-mass velocity & $\gamma_{\rm Be}$\,[km\,s$^{-1}$] & $10.5\pm 0.3$ \\ 
B star eccentricity & $e_{\rm B}$ & $<0.037$  \\
Be star eccentricity & $e_{\rm Be}$ & $<0.013$  \\
B star RV scatter & $s_{\rm B}$ [km\,s$^{-1}$] & $4\pm 0.5$ \\
Be star RV scatter & $s_{\rm Be}$ [km\,s$^{-1}$] & $<1.0$ \\
Mass ratio  &  $q = M_{\rm B}/M_{\rm Be}$  & $0.071\pm 0.032$ \\
Orbital inclination & $i\,[\rm deg]$ & $32.1^{+3.0}_{-2.6}$  \\
Separation  & $a$ [R$_{\odot}$]   &  $96\pm 8$ \\
\hline

\multicolumn{3}{l}{\bf{B star abundances relative to Solar}}  \\ 
Helium & [He/H]  & $0.55 \pm 0.2$ \\
Carbon  & [C/H] & $-0.5 \pm 0.2 $ \\
Nitrogen  & [N/H]  & $1.1 \pm 0.3 $ \\
Oxygen & [O/H] & $0.3 \pm 0.3 $ \\
Neon  & [Ne/H] & $0.15 \pm 0.2  $ \\
Magnesium  & [Mg/H] & $0.2 \pm 0.2 $ \\
Aluminum  & [Al/H] & $0.4 \pm 0.2 $ \\
Silicon  & [Si/H] & $0.3 \pm 0.2 $ \\
Sulfur  & [S/H] & $-0.3 \pm 0.2 $ \\
Argon  & [Ar/H] & $0.4 \pm 0.2 $ \\
Iron & [Fe/H]  & $0.2 \pm 0.2$ \\

\hline
\multicolumn{3}{l}{\bf{Be star disk parameters}} \\
Projected outer disk rotation velocity  &  $v_{\rm out}\sin i\,[\rm km\,s^{-1}]$  & $105 \pm 4$ \\
Ratio of inner to outer disk radii & $R_{\rm inner}/R_{\rm outer}$ & $0.24 \pm 0.05$ \\
Fe II emissivity exponent & $\alpha=-{\rm d}\ln j/{\rm d}\ln r$ & $1.6\pm 0.3$ \\
Outer disk radius & $R_{\rm outer}\,[R_{\odot}]$ & $33^{+5}_{-4} $ \\
Inner disk radius & $R_{\rm inner}\,[R_{\odot}]$ & $7.8^{+2.1}_{-1.8} $ \\

\hline
\end{tabular}
\begin{flushleft}

\label{tab:system}
\end{flushleft}
\end{table}

\begin{figure}
\includegraphics[width=\columnwidth]{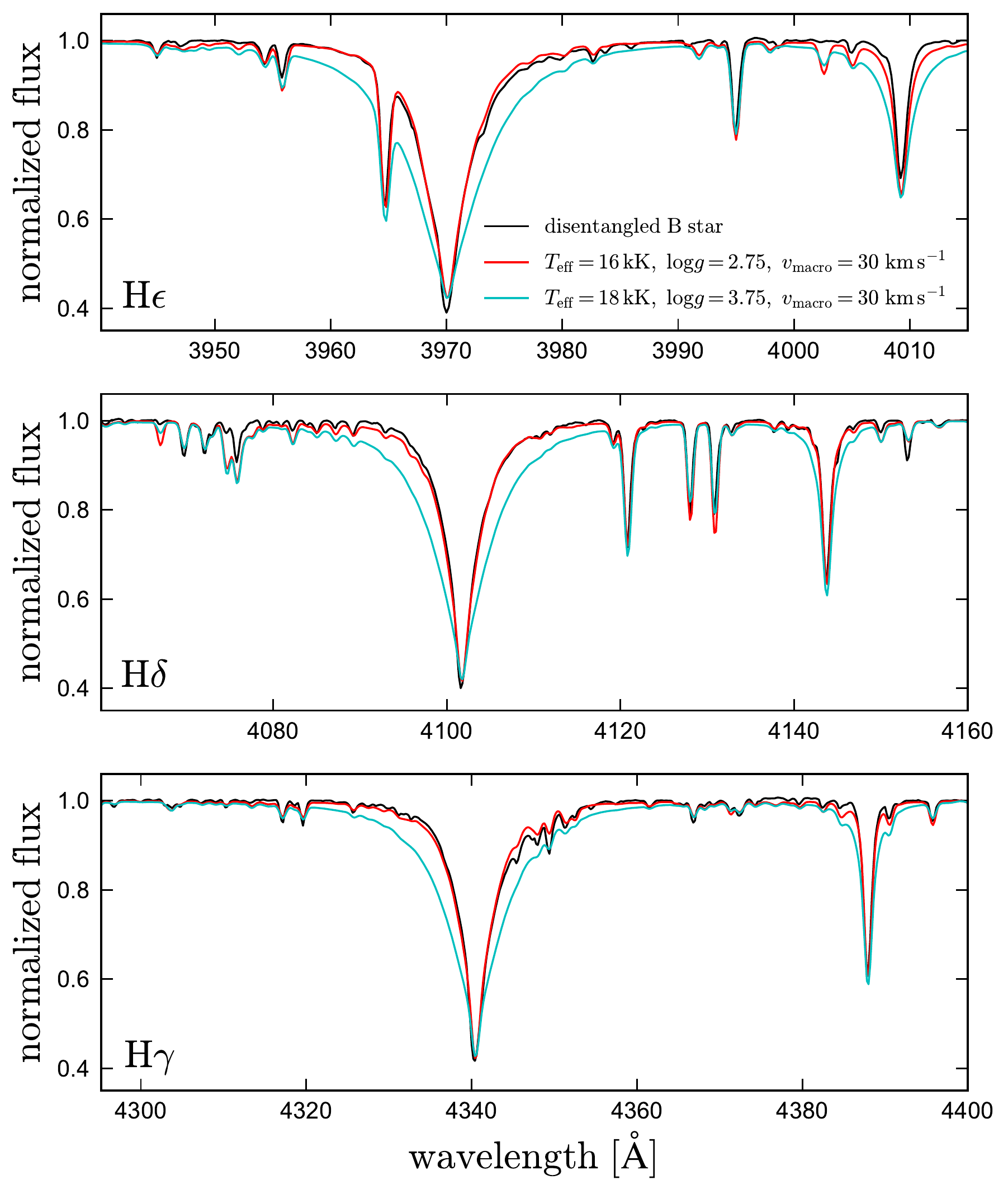}
\caption{Balmer lines as a surface gravity diagnostic. Black lines show the spectrum of the B star obtained through spectral disentangling. Red lines shows a TLUSTY model spectrum with $\log g = 2.75$, which reproduces the observed Balmer line profiles reasonably well. Cyan lines show model spectra with $\log g = 3.75$, as expected for a normal B star near the end of the main sequence. For this value of $\log g$, the wings of the Balmer and He I lines are much too wide to match the observed spectrum. The model $T_{\rm eff}$ values are chosen to match the observed ionization ratios at a given $\log g$ (Figure~\ref{fig:Teff}). For other choices of  $T_{\rm eff}$, the disagreement between the $\log g = 3.75$ model and the observed spectrum is even more severe.}
\label{fig:balmer}
\end{figure}

We fit the disentangled spectra of both components using 1D-NLTE model spectra computed with TLUSTY \citep{Hubeny_1995, Hubeny_2017}. We use the BSTAR06 grid, which has $\Delta T_{\rm eff} = 1\,\rm kK$ and $\Delta \log g = 0.25\,\rm dex$, \citep{Lanz_2007} in an initial grid search and then use the radiative transfer code SYNSPEC \citep{Hubeny_2011} to generate spectra with finer grid spacing, as well as spectra with different abundance patterns than assumed in the BSTAR06 grid (Section~\ref{sec:abundances}).  We constructed grids of spectra with 4 values of the microturbulent velocity: $v_{\rm mic} = (2,5,10,15)\,\rm km^{-1}$. We find $v_{\rm mic}=10\,\rm km\,s^{-1}$ to provide the best fit and adopt this value in the rest of our analysis \citep[see also][]{Bodensteiner_2020}, but we account for a $\pm 5\,\rm km\,s^{-1}$ uncertainty in $v_{\rm mic}$ when calculating uncertainty in other spectral parameters and abundances. 

\begin{figure*}
    \includegraphics[width=\textwidth]{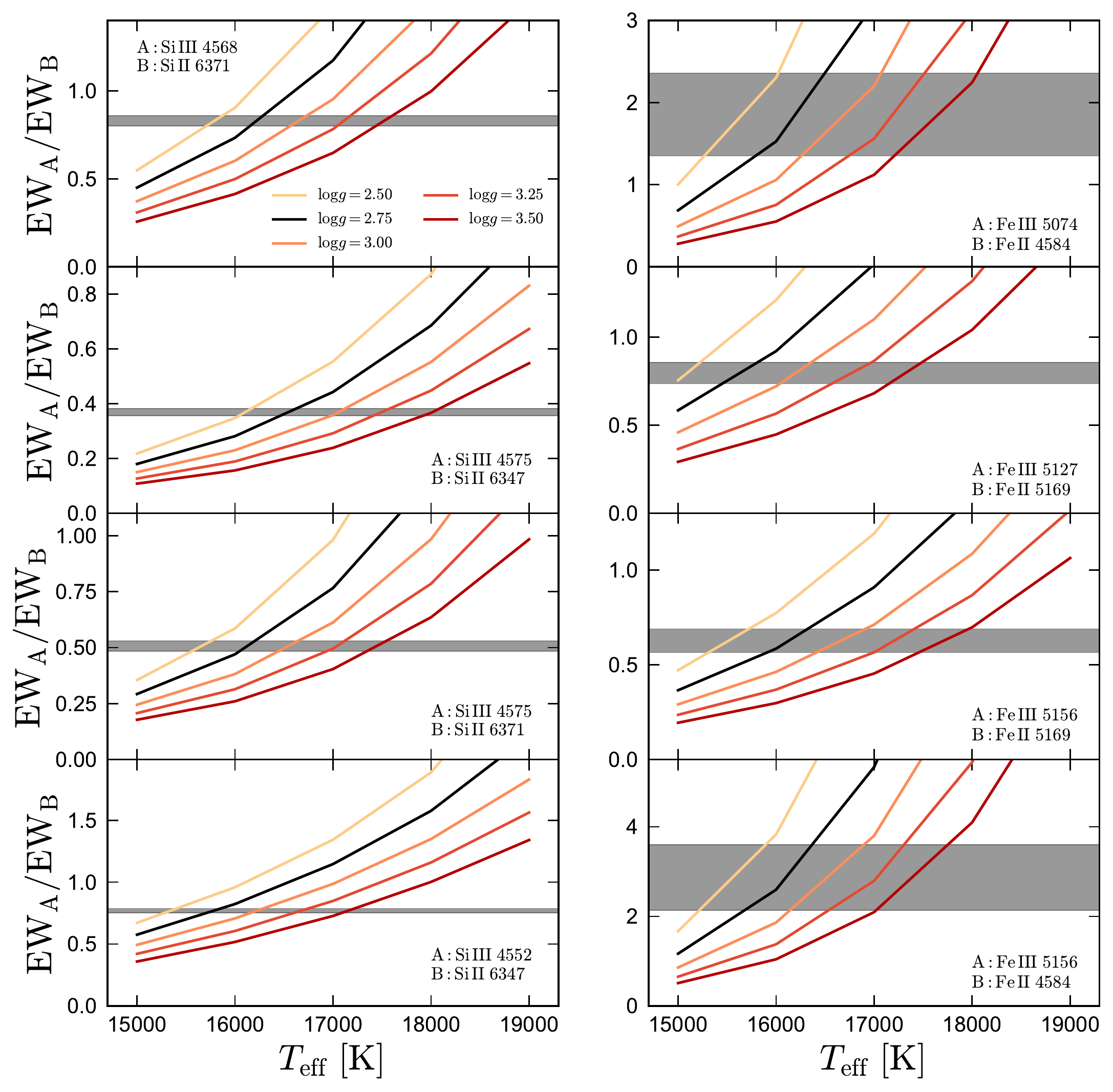}
    \caption{Temperature of the B star from ionization equilibrium. Shaded regions show the observed ratios of the equivalent widths of Si III and Si II lines (left columns) and Fe III and Fe II lines (right columns). Lines show the ratios predicted for models of different $\log g$ as a function of $T_{\rm eff}$. Given the best-fit $\log g\approx 2.75$ from the Balmer lines (Figure~\ref{fig:balmer}), the best-fit $T_{\rm eff}$ lies between 15500 and 16500\,K for all line ratios.}
    \label{fig:Teff}
\end{figure*}

We perform an iterative fit for the temperature and gravity of the B star, first measuring $T_{\rm eff}$ based on equivalent width ratios of lines in different ionization states (Figure~\ref{fig:Teff}), then measuring $\log g$ based on the wings of broad lines, then refining the $T_{\rm eff}$ estimate taking the estimated $\log g$ into account, and so forth. Detailed abundances are measured in a final step, once a converged solution for $T_{\rm eff}$ and $\log g$ is found.
To estimate systematics due to model uncertainties, we also compared to 1D-LTE model spectra generated with SYNTHE \citep{Kurucz_1993} from ATLAS-12 model atmospheres \citep{Kurucz_1970, Kurucz_1979, Kurucz_1992}. 
We find consistent solutions for the two sets of model spectra within 1\,kK in $T_{\rm eff}$ and 0.2\,dex in $\log g$. 

The model spectra used in determining the temperature and gravity of the B star assume a solar abundance pattern. We show in Section~\ref{sec:abundances} that the B star is enhanced in helium. We have checked that this does not significantly change our inferred atmospheric parameters, and the level of helium enhancement we find has a minimal effect on the shapes of the Balmer lines.

Our final best-fit parameters for both stars are listed in Table~\ref{tab:system}. The reported uncertainties in atmospheric parameters include formal fitting uncertainties due to noise in the spectra, uncertainty in the flux ratio, systematic differences between fits with TLUSTY/SYNSPEC and Kurucz spectra, and covariances between atmospheric parameters. Systematic uncertainty dominates the error in $T_{\rm eff}$, and because $T_{\rm eff}$ and $\log g$ are strongly covariant, also contributes significantly to the uncertainty in $\log g$.

A critical result of our analysis is that the surface gravity of the B star is lower than that of a main-sequence star, $\log g \approx 2.75$. The main constraint on $\log g$ comes from the wings of the Balmer lines and some He I lines, which become broader with increasing $\log g$. This is illustrated in Figure~\ref{fig:balmer}, which shows a zoom-in on the Balmer lines H$\epsilon$, H$\delta$, and H$\gamma$ in the disentangled spectrum of the B star. We compare the best-fit model, which has $\log g = 2.75$, to one with $\log g = 3.75$, the approximate value assumed by \citealt{Rivinius_2020}, and the best-fit $T_{\rm eff}$ for that $\log g$. The $\log g = 3.75$ model predicts much wider Balmer lines than are found in the observed spectrum. The poor fit of the higher $\log g$ model is also evident in the wings of the 4009\,\textup{\AA} and 4388\,\textup{\AA} He I lines. We note that while the superiority of the $\log g = 2.75$ model is clear in Figure~\ref{fig:balmer}, this is only the case because we have disentangled the spectra of the two components. In the composite spectra, the Balmer and He I lines appear much broader due to the broader absorption lines of the Be star.

The effective temperature of the B star is constrained  from equivalent width ratios of lines of the same element in different ionization states. Higher ionization states become increasing populated as temperature increases. The ratios of e.g. Si III to Si II equivalent widths are therefore sensitive to $T_{\rm eff}$ and relatively insensitive to the absolute Si abundance. Unlike absolute abundance measurements, equivalent width ratios are also insensitive to the assumed luminosity ratio of the two components, since light from the companion dilutes all lines by a similar factor. 

Figure~\ref{fig:Teff} compares the observed equivalent width ratios of Si III/II and Fe III/II lines to predictions from TLUSTY models for a range of $T_{\rm eff}$ and $\log g$. Our best-fit model with $\log g=2.75$ is shown in black; colored lines show other values of $\log g$. All the line ratios we consider imply a temperature of $T_{\rm eff} = (16\pm 0.5)\,\rm kK$ for $\log g =2.75$, with values that are respectively higher and lower for higher or lower $\log g$. 

Rapid rotation washes out most of the Be star's lines, making it infeasible to measure $T_{\rm eff}$ from ionization state ratios. The constraint on $T_{\rm eff}$ comes primarily from the Balmer lines and the strength of Si II and He I lines. The star's gravity can still be constrained from the shape of the Balmer lines, which are sufficiently broad that their wings are not strongly affected by rotation.

A natural concern when working with echelle spectra is that the profiles of broad lines such as the Balmer lines could be distorted by the blaze function and/or continuum normalization process, leading to a bias in $\log g$. To assess the reliability of our derived $T_{\rm eff}$ and $\log g$, we analyzed an archival FEROS spectrum of the slowly rotating standard B star HR 5285 that was presented and analyzed by \citet{Nieva_2007}. The spectral resolution and SNR of this spectrum is similar to that of the HR 6819 data. We apply the same continuum normalization routine and procedure for estimating atmospheric parameters used for HR 6819 to this spectrum and obtain $T_{\rm eff}= 21\,\rm kK$ and $\log g = 4.25$, in good agreement with the literature values (\citealt{Vrancken_1996} found $T_{\rm eff}=22.6\pm 0.9\,\rm kK$ and $\log g = 4.2\pm 0.1$, while \citealt{Nieva_2012} found $T_{\rm eff}=20.8\pm 0.3\,\rm kK$ and $\log g = 4.22\pm 0.05$). This allays any fear that our inferred $\log g$ values are significantly biased. 

\subsection{Radial velocities}
\label{sec:velocites}

We measure radial velocities for both components at each epoch by fitting the composite spectra as a sum of the two disentangled spectra obtained in Section~\ref{sec:disentangle}, with the radial velocities of both components left free.  We also leave the luminosity ratio free, because at least one component of the system is photometrically variable at the 5-10\% level \citep{Rivinius_2020}. We mask regions of the spectrum containing diffuse interstellar bands, telluric absorption, and a few lines that are poorly fitted. 

We divide the spectrum into four segments, fitting the RVs in each segment independently. This allows us to empirically estimate RV uncertainties from the scatter between different segments. It also allows us to test whether the RVs of the emission lines, which trace the Be star's disk and circumstellar envelope, vary coherently with the absorption lines, which trace the photosphere. In particular, the region of the spectrum with $5100 < \lambda/\textup{\AA} < 5400$ contains no strong absorption lines from the Be star but contains seven Fe II emission lines, which likely originate in a disk around the Be star.

As discussed by \citet{ElBadry_2020} and \citet{AbdulMasih_2020}, RVs measured from emission lines can be biased if contamination from absorption lines is not taken into account. Our procedure for measuring RVs is not susceptible to the bias pointed out in these works, because the spectra of both components are modeled simultaneously. 

\begin{figure}

    \includegraphics[width=\columnwidth]{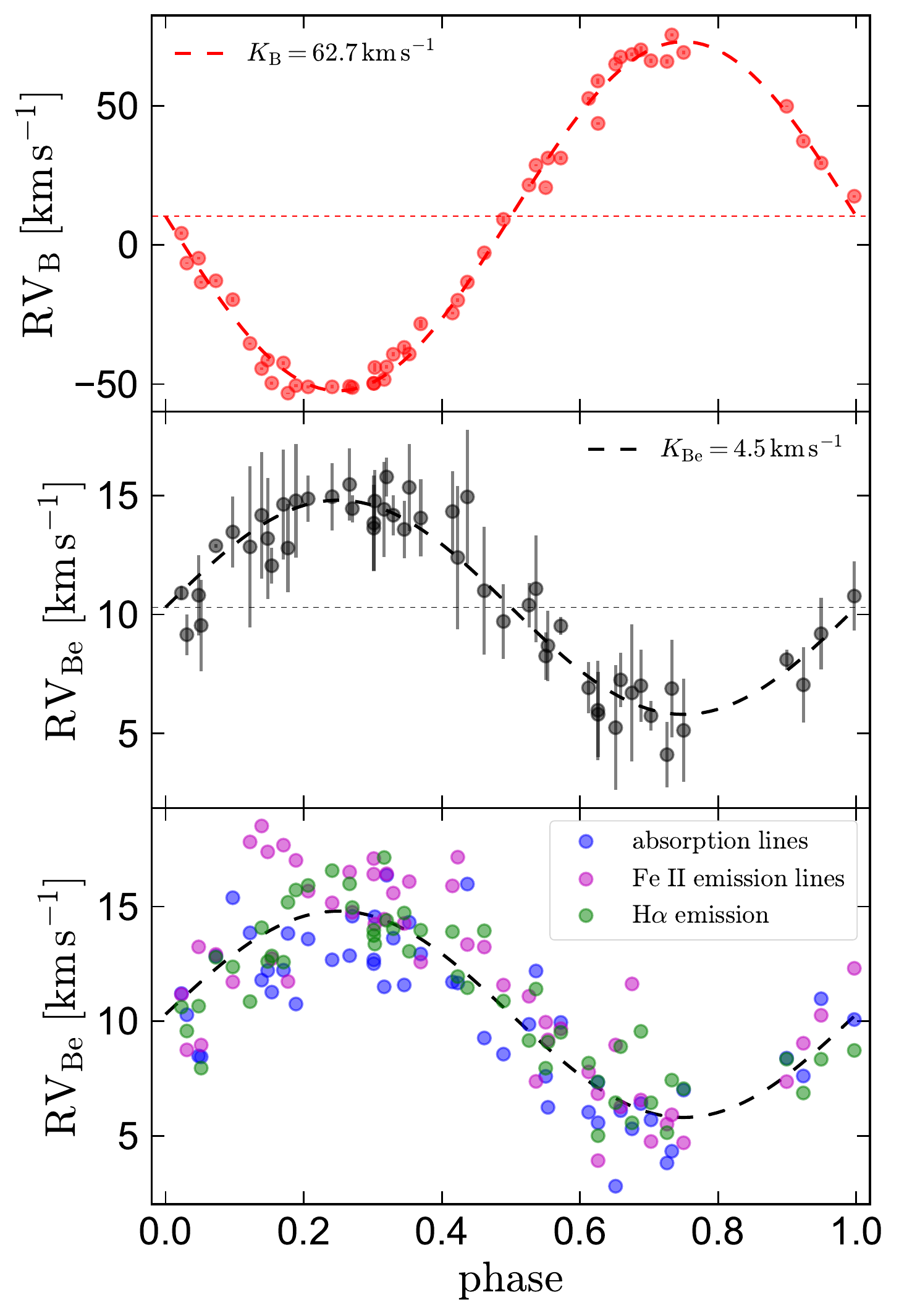}
    \caption{Radial velocities for the B star (top panel) and Be star (middle and bottom panels), phased to a period of 40.3 days. RVs for both components are measured simultaneously by fitting the composite spectra with the single-star templates obtained from spectral disentangling. The Be star exhibits sinusoidal RV variability in anti-phase with the B star, suggesting the two components are orbiting each other with a period of 40.3 days. The much lower velocity amplitude of the Be star implies a mass ratio $M_{\rm B}/M_{\rm Be}=0.071\pm 0.032$. The bottom panel compares velocities for the Be component measured from absorption lines (presumably originating in the stellar photosphere) and emission lines (presumably originating in a circumstellar disk). Emission and absorption line RVs vary coherently. }
    \label{fig:velocities}
\end{figure}

Figure~\ref{fig:velocities} shows the measured epoch RVs for both components, phased to a period of 40.3 days. Error bars in the top two panels show the scatter between RVs fit to different spectral regions. The Be star displays sinusoidal RV variability with the same period as the B star, suggesting that they are orbiting each other. Emission and absorption lines show consistent variability, implying that the Be star and its disk move coherently. This implies that the disk is likely circumstellar, not circumbinary.  For the B star, the RV uncertainties are smaller than the symbols for most points, suggesting that there is intrinsic RV scatter. 

We fit the RVs of both components independently with Keplerian orbits, without the requirement that they have the same center-of-mass velocity or phase. Along with the 6 standard Keplerian orbital parameters, we fit for an intrinsic RV scatter or ``jitter'' term (see \citealt{Elbadry_2018}, their Equation 9), which can represent either intrinsic RV variability not captured in the model or underestimated uncertainties. We initially search for the best-fit orbital solution using simulated annealing, and then sample from the posterior in the vicinity of this solution using a Markov chain Monte Carlo method, as described in \citet{Elbadry_2018}. Given the large number of RV measurements (51), there is little danger of the sampler getting stuck in a local minimum. Our best-fit orbital parameters are listed in Table~\ref{tab:system}.

Our best-fit velocity amplitude for the B star is $K_{\rm B} = 62.7\pm 1.0 \,\rm km\,s^{-1}$, which is consistent with the value measured by \citet{Rivinius_2020}. Our constraints on $P$ is also consistent, but our uncertainties are larger, because we did not include the additional RVs from 1999 in our fit.  Unlike \citet{Rivinius_2020}, we do not measure any significant eccentricity: we find an upper limit of $e < 0.037$ for the B star, but the orbit is consistent with $e=0$ once RV scatter is accounted for. For the Be star, we measure $e < 0.13$, again consistent with a circular orbit. The formal uncertainty in $K_{\rm Be}$ obtained from fitting the RVs shown in Figure~\ref{fig:velocities} is only $0.3\,\rm km\,s^{-1}$, but the true uncertainty is larger, since the points shown in Figure~\ref{fig:velocities} were obtained after fixing $K_{\rm Be}=4.5\,\rm km\,s^{-1}$ in spectral disentangling (see Appendix~\ref{sec:disentangle_details}). 

For the B star, we find an intrinsic RV scatter of $\approx 4\,\rm km\,s^{-1}$. We did not find evidence of additional periodicity in the residuals. As we show in Appendix~\ref{sec:variability}, the B star's RV scatter is accompanied by modest variability in its spectral line profiles. HR 6819 also exhibits photometric variability with brightness variations of up to five percent in the total light (implying variations of ~10\% if the light comes primarily from one component) on timescales down to half a day \citep{Rivinius_2020}. The most likely explanation is that the B star is a slowly pulsating B star (SPB star; e.g. \citealt{Waelkens_1991}) undergoing non-radial $g$-mode pulsations that cause the RV scatter, line profile variability, and photometric variability. The B star's atmospheric parameters, the amplitude of the RV and photometric variability, and the variability timescale are all within the normal range found for SPB stars \citep[e.g.][]{Dziembowski_1993}. We note, however, that photometric variability is also common for Be stars, due to both pulsations and variations in the structure of the disk \citep{Rivinius_2013}, so some of the observed photometric variability could also come from the Be star.

If the masses of both components are known, the inclination of the binary's orbital plane can be measured from the velocity semi-amplitudes. For a circular orbit, 
\begin{align}
    \label{eq:sini}
    \sin^{3}i=\frac{\left(1+K_{{\rm Be}}/K_{{\rm B}}\right)^{2}}{2\pi G}\frac{K_{{\rm B}}^{3}P}{M_{{\rm Be}}}.
\end{align}
Taking $M_{\rm Be}=6.7^{+1.9}_{-1.5} M_{\odot}$, a constraint which we obtain by comparing the Be star's temperature and surface gravity to MIST evolutionary tracks \citep[see Appendix~\ref{sec:distance_anchor};][]{Dotter_2016, Choi_2016}, we obtain obtain $i=32.1^{+3.0}_{-2.6}$\,deg. The mass ratio of the two components is constrained from the relative velocity amplitudes to $M_{\rm B}/M_{\rm Be}=0.071\pm 0.032$. This leads to $M_{\rm B} = 0.47^{+0.28}_{-0.22} M_{\odot}$, where the dynamical mass ratio and the mass of the Be star contribute roughly equally to the uncertainty. 

\subsection{Luminosity}
\label{sec:lum}

If the distance to HR 6819 is known, then its absolute magnitude can distinguish between models in which the B star is a normal star and models in which it was recently stripped. The bolometric luminosity of a star can be approximated as 
\begin{align}
    \label{eq:L}
    \frac{L}{L_{\odot}}&=\left(\frac{R}{R_{\odot}}\right)^{2}\left(\frac{T_{{\rm eff}}}{T_{{\rm eff,\odot}}}\right)^{4}=\left(\frac{M}{M_{\odot}}\right)\left(\frac{g}{g_{\odot}}\right)^{-1}\left(\frac{T_{{\rm eff}}}{T_{{\rm eff,\odot}}}\right)^{4}\\
    \label{eq:L2}
    &=1400\left(\frac{M}{0.5M_{\odot}}\right)\left(\frac{10^{\log g}}{10^{2.75}}\right)^{-1}\left(\frac{T_{{\rm eff}}}{16\,{\rm kK}}\right)^{4}.
\end{align}
As discussed by \citet{Rivinius_2020}, a normal star with the temperature and gravity of the B star in HR 6819 would have a minimum mass of at least 5\,$M_{\odot}$. It would thus be a minimum of 10 times brighter than a $0.5 M_{\odot}$ stripped star with the same $T_{\rm eff}$ and $\log g$. And as we discuss below, the minimum mass for a normal B star matching HR 6918's spectral type is even higher, of order $10\,M_{\odot}$, once the lower $\log g $ measured from the disentangled spectrum is taken into account. 

\subsubsection{Distance}
\label{sec:distance}
{\it Gaia} measured a parallax of $\varpi = 2.71\pm 0.12$\,mas for HR 6819 ({\it Gaia} DR3 \texttt{source\_id} 6649357561810851328), implying a distance $d\approx 369\pm 17\,\rm pc$ \citep{Gaia2020}.\footnote{The {\it Gaia} DR2 parallax was $\varpi = 2.91\pm 0.18$\,mas. The distance errors implied by the parallax are slightly asymmetric, $d=369^{+17}_{-16}$ pc; we symmetrize them for simplicity. } At this distance, the projected semimajor axis would be $\theta \approx 1.2$\,mas. Since this is almost half the parallax, and the astrometric solution did not account for orbital motions, the parallax uncertainty is likely underestimated. The re-normalized unit weight error for the source is large, \texttt{ruwe} = 1.45, also indicating a problematic astrometric solution \citep[e.g.][]{Fabricius2020}. Given that the {\it Gaia} data span many orbital periods, the {\it Gaia} distance is not expected to be catastrophically biased, but the parallax uncertainty should not be taken at face value.

\citet{Rivinius_2020} derived an independent estimate of the distance to HR 6819 by comparing its flux-calibrated UV spectrum to models with temperature and radius matching their fit to the spectrum. They estimated $d=(310\pm 60)\,\rm pc$, which is also consistent with the {\it Gaia} distance. The largest source of uncertainty was the unknown luminosity ratio between the two components. %We adopt this same estimate in our analysis. 
Although the spectral type we infer for the B star is significantly different from that estimated by \citet{Rivinius_2020}, only the temperature and radius are relevant for the distance estimate. The values they adopted ($T_{\rm eff}= 16\,\rm kK$ and $R=(5.5\pm 0.5) R_{\odot}$) are consistent with our best-fit values, because our lower mass estimate is almost exactly compensated for by our lower estimate of $\log g$.  

\citet{Elbadry2021} found that bright sources with \texttt{ruwe} > 1.4 typically have parallax uncertainties underestimated by up to a factor of 2. We adopt the {\it Gaia} distance estimate in our analysis, but we conservatively inflate the parallax uncertainty by a factor of $\approx$2 due to the large \texttt{ruwe} and expected orbital motion. We assume $d=369\pm 34\,\rm pc$. To account for the possibility of more severe problems with the {\it Gaia} distance, we also present stellar parameters that are inferred without taking the distance into account.

\subsubsection{Synthetic photometry}
We used synthetic photometry from MIST isochrones \citep{Dotter_2016, Choi_2016} to place models for HR 6819 on the color-magnitude diagram (CMD), assuming solar metallicity. 
We assume an extinction of $E(B-V)=0.135$, as reported by \citet{Wegner_2002}, and a \citet{Cardelli_1989} extinction curve with $R_V=3.1$. The SFD map estimates $E(B-V)=0.13$ at the position HR 6819 \citep{Schlegel_1998}.

\begin{figure*}
    \includegraphics[width=\textwidth]{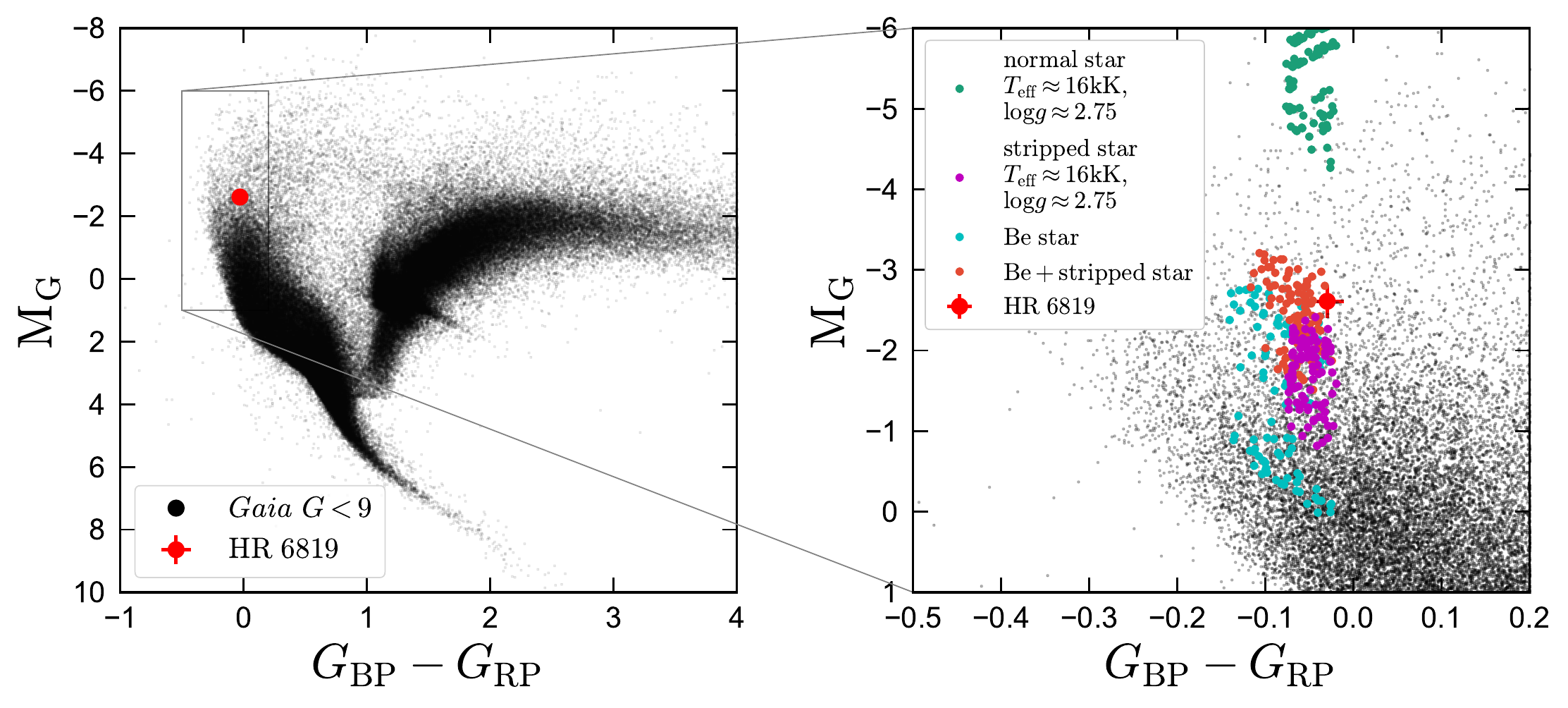}
    \caption{Color-magnitude diagram. Left panel compares HR 6819 to all {\it Gaia} sources brighter than $G=9$. Right panel compares it to synthetic photometry from isochrones. Green points show predictions for a normal star, near the end of the main sequence, with temperature and surface gravity consistent with HR 6819. Such a star would be much brighter than HR 6819 and is ruled out. Magenta points show the CMD position of a stripped star with the same temperature and gravity. Due to its lower mass and smaller radius, a stripped star is much fainter. Cyan points show models with temperature and gravity consistent with our measurements for the Be star. Orange points show unresolved binaries composed of a Be star and stripped star model; these are consistent with the observed color and magnitude of HR 6819. }
    \label{fig:cmd}
\end{figure*}

We began by searching the MIST model grid for ``normal'' stellar models with atmospheric parameters similar to the B star ($15 < T_{\rm eff}/{\rm kK} < 17$ and $2.4 < \log g < 3.1$). These models have masses ranging from $9 M_{\odot}$ to as much as $20 M_{\odot}$ (see Figure~\ref{fig:kiel_diag}). This is significantly higher than the mass of $(5-6) M_{\odot}$ that would be inferred if the B star had $\log g = 3.5-4$, as assumed by \citealt{Rivinius_2020}. These models are shown in green in the right panel of Figure~\ref{fig:cmd}. Even without the contribution from the Be star, these models are far brighter than HR 6819. 

We next calculate synthetic photometry for a stripped star with the same range of $T_{\rm eff}$ and $\log g$ as the models described above, accounting for the lower mass using Equation~\ref{eq:L2}. These models are shown in magenta in Figure~\ref{fig:cmd}; we assume masses between 0.4 and 0.6 $M_{\odot}$ for the stripped star. We also take synthetic photometry for the Be star from models with $15 < T_{\rm eff}/{\rm kK} < 20$ and $3.5 < \log g < 4.25$. These models are shown in cyan; we do not attempt to correct for gravity darkening due the the Be star's rapid rotation. Finally, we calculate the combined magnitude and color of randomly paired Be star and stripped star models (orange points). These are generally consistent with the color and magnitude of HR 6819. 

This analysis reveals that the B star in HR 6819 cannot be a normal B supergiant -- its luminosity is too low. A consistent solution can, however, be obtained if the B star has unusually low mass, of order $0.5 M_{\odot}$.
The poor agreement between a ``normal B star'' model and the observed luminosity of HR 6819 cannot be alleviated by simply assuming a different distance. A distance of 3-5 times our fiducial value would make the observed magnitude of HR 6819 compatible with a normal B star (green points in Figure~\ref{fig:cmd}). But in this case, the B star would be a factor of $\sim 20$ brighter than the Be star, violating the observational constraint that the two components are roughly equally bright. 

\subsubsection{Luminosities, masses and radii}
\label{sec:masses_radii}
We estimate the mass and radius of the Be star by comparing its temperature and surface gravity to a grid of MIST evolutionary tracks (Figure~\ref{fig:kiel_diag}). We then calculate the mass of the B star from the measured mass ratio, and its radius from this mass and its measured surface gravity. The bolometric luminosity of each component is then calculated through Equation~\ref{eq:L}. The resulting stellar parameters are reported in Table~\ref{tab:system}. 

In Appendix~\ref{sec:distance_anchor}, we calculate an estimate of the same stellar parameters that is anchored on the distance to HR 6819 rather than on evolutionary tracks. We obtain consistent solutions between the two methods.

\subsection{Emission lines}
\label{sec:emission}

\begin{figure*}
    \includegraphics[width=\textwidth]{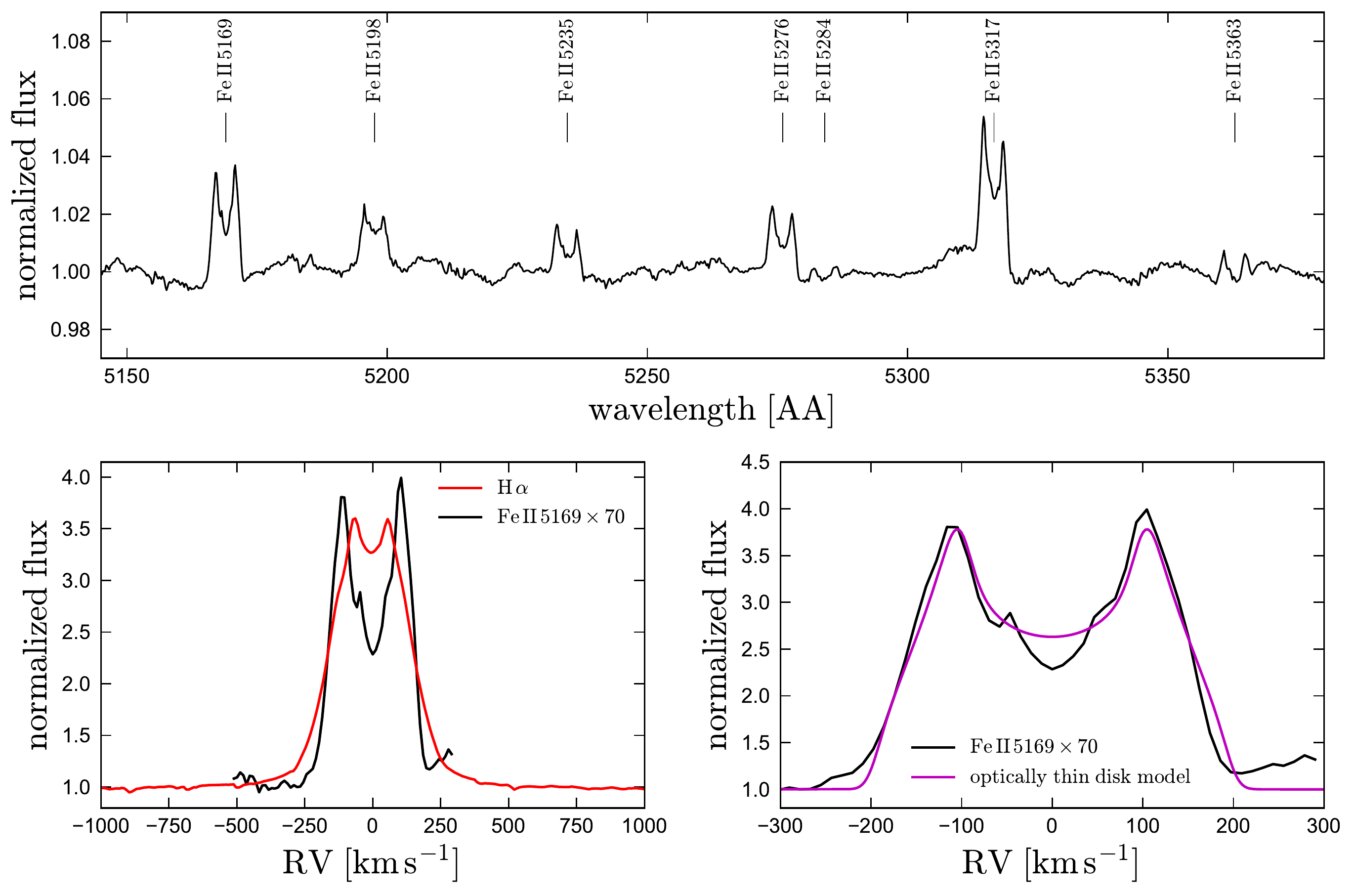}
    \caption{Line emission from the disk and circumstellar envelope of the Be star. Top panel shows a section of the disentangled Be star's spectrum that displays several Fe II emission lines. Each line is double-peaked, as is expected for an optically thin Keplerian disk. Bottom left panel compares the line shapes of H$\alpha$ and one of the Fe II lines. The H$\alpha$ line has broader wings and a narrower peak, likely due to its higher optical depth. Bottom right panel compares the shape of the Fe II line to a model fit assuming a truncated, optically thin disk (see text). The best-fit model has an inner radius $R_{\rm inner}\approx 8.6 R_{\odot}$ and outer radius $R_{\rm outer} \approx 35 R_{\odot}$, reasonably consistent with the observed stellar radius and orbital semi-major axis. }
    \label{fig:emission_lines}
\end{figure*}

As is often the case in classical Be stars, the disentangled Be star spectrum contains clearly double-peaked emission in the Balmer lines and in many Fe II lines. Weak double-peaked emission lines are also found in some C II and Si II lines. Figure~\ref{fig:emission_lines} highlights the H$\alpha$ line (which is by far the strongest of the star's emission lines) and several of the Fe II lines. We correct the emission lines for contamination from the Be star's absorption lines by subtracting a TLUSTY model spectrum with $T_{\rm eff} = 18\,\rm kK$, $\log g=3.75$, and $v\,\sin i =180\,\rm km\,s^{-1}$. This has little effect on the Fe II lines, but it increases the amplitude and slightly changes the shape of the H$\alpha$ line.  

The integrated emission line profile of an optically thin, kinematically cold Keplerian disk with known emissivity profile can be calculated analytically \citep[e.g.][]{Smak_1969}. Following \citet{Shen_2019}, we assume that the emissivity profile can be approximated by a power law, $j(r) \propto r^{-\alpha}$, and that the emissivity is sharply truncated at inner radius $r = R_{\rm inner}$ and outer radius $R_{\rm outer}$. This truncation could either represent the geometric edge of the disk, or a change in the ionization state. We assume an intrinsic thermal broadening with isothermal sound speed $c_s = 10\,\rm km\,s^{-1}$, appropriate for ionized gas with temperature near $10^4$\,K.

Given these assumptions, we fit the emission line profile of the Fe\,II 5169 line. The free parameters of the fit are the projected rotation velocity at the outer edge of the disk, $v_{{\rm out}}\sin i=(GM_{\rm Be}/R_{{\rm outer}})^{1/2}\sin i$, $r_{\rm inner} = R_{\rm inner}/R_{\rm outer}$, the ratio of the inner disk edge to the outer disk edge, and the emissivity exponent $\alpha$. We obtain $\alpha \approx 1.6$, $v_{\rm out} \approx 105\,\rm km\,s^{-1}$, and $r_{\rm inner} \approx 0.24$ (see Table~\ref{tab:system}). This leads to  $R_{\rm inner}\approx 7.8 R_{\odot}$ and $R_{\rm outer} \approx 33 R_{\odot}$. These parameters can be interpreted straightforwardly from the observed line profiles: $R_{\rm inner}$ is constrained by the maximum width of the emission lines (since the largest Keplerian velocities are found at the inner edge of the disk) and $R_{\rm outer}$ is constrained by the RV separation of the two peaks (since the total emission is maximized at the outer edge of the disk). 

We note that applying the same fitting to the H${\alpha}$ line yields significantly different results, including an unphysically small $R_{\rm inner}$. This is because repeated absorption and re-emission of H$\alpha$ photons, as well as electron scattering,  significantly broadens the wings of the H$\alpha$ line. This makes it an unreliable kinematic tracer \citep[e.g.][]{Hummel_1992}. Fe II lines have lower optical depth and are therefore more reliable optical tracers of the disk \citep{Hanuschik_1988, Dachs_1992}, though they too many not be completely optically thin in the inner parts of the disk \citep{Arias_2006}.

For our best-fit orbital parameters, the binary's semimajor axis is $a\approx 0.44\,\rm AU \approx 96\,R_{\odot}$. That is, the disk emission extends to about $0.34a$. For a binary of mass ratio $0.07$, the largest stable streamline at which a disk can be maintained is $r_{\rm max}\approx 0.44a$ \citep{Paczynski_1977}, so the disk likely extends to near its maximum stable radius. 

\subsection{Chemical abundances}
\label{sec:abundances}
We use SYNSPEC \citep{Hubeny_2011} to generate grids of model spectra with $T_{\rm eff}= 16\,\rm kK$ and $\log g = 2.75$, varying each of the 11 elements listed in Table~\ref{tab:system} with a grid spacing of 0.1\,dex. All abundances are measured relative to the Solar abundance pattern, which is taken from \citet{Grevesse_1998}.  The atmospheric structure is taken from the BSTAR06 atmosphere grid, which was generated with TLUSTY assuming $Z=0.02$ and the Solar abundance pattern. We assume a microturbulent velocity $v_{\rm mic}=10\,\rm km\,s^{-1}$ but also experimented with varying $T_{\rm eff}$, $\log g$, and $v_{\rm mic}$ within their uncertainties.

Unlike the line ratios shown in Figure~\ref{fig:Teff}, the abundances are sensitive to the luminosity ratio of the two components. The smaller the ratio $L_{\rm B}/L_{\rm Be}$, the deeper the lines of the B star, and the larger the inferred abundances. Our reported abundance uncertainties account for a $\pm 5\%$ uncertainty in the luminosity ratio. Uncertainties in $T_{\rm eff}$, $v_{\rm mic}$, and the flux ratio, not the noise in the spectrum, dominate the abundance uncertainties.

The most striking result of our abundance analysis is that the surface of the B star is significantly enriched in nitrogen and helium, and significantly depleted of carbon. This suggests that material on the B star's surface has been processed by the CNO cycle, as has also been reported for LB-1 \citep{Irrgang_2020, Shenar_2020}. An enhancement of nitrogen and helium at the expense of carbon and oxygen is expected in the stripped star scenario: the material at the surface of the star today was inside the convective core of the progenitor when it was on the main sequence and is thus contaminated with fusion products. 

Unlike in LB-1, we do not find a significant depletion of oxygen for the B star. Most other elements are consistent with, or slightly enhanced relative to, the Solar value.  We note that \citet{Irrgang_2020} found most metals to be depleted in LB-1, but this is likely at least in part due to the unaccounted-for line dilution by the Be star, which makes all lines appear weaker by roughly a factor of 2. 

We do not attempt to measure detailed abundance for the Be star, since most metal lines are washed out by rotation, uncertainties in continuum normalization will significantly affect the interpretation of weak and highly broadened lines, and contamination by disk emission complicates the interpretation of absorption line equivalent widths. Besides H and He, the strongest lines visible in the Be star spectrum come from C II and Si II. The strongest tension between the disentangled Be star spectrum and the TLUSTY model occurs for the $4075\,\textup{\AA}$ CII line, which is much weaker than predicted. This suggests that the Be star's surface is also depleted of C, which is consistent with our expectations if the Be star recently accreted much of the CNO-processed envelope of the B star. 

We defer more detailed analysis and modeling of HR 6819's abundance pattern to future work. Given the B star's unusually large surface He abundance, it would be useful to recalculate the atmospheric structure while taking the abundance variations into account.

\subsection{Evolutionary history}
\label{sec:evol}

To better understand the possible formation pathways and future evolution of HR 6819, we began by searching the BPASS (v2.2; \citealt{Eldridge_2017}) library of binary evolution models for models that go through a phase similar to the current state of HR 6819. We then investigated a few possible formation histories in more detail using Modules for Experiments in Stellar Astrophysics \citep[MESA, version 12778;][]{Paxton_2011, Paxton_2013, Paxton_2015, Paxton_2018, Paxton_2019}.

\begin{figure}
    \centering
    \includegraphics[width=\columnwidth]{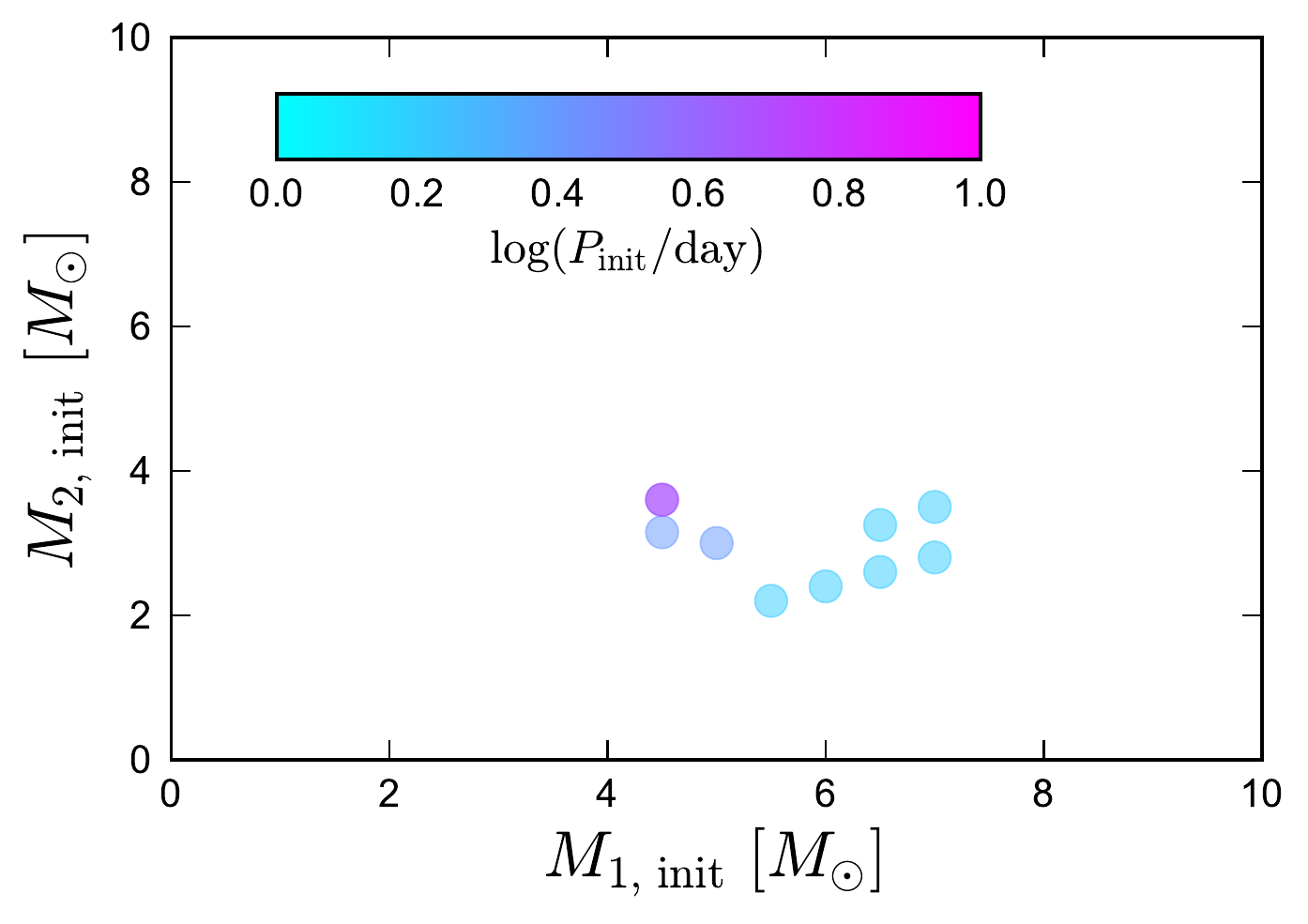}
    \caption{Initial masses and periods of models in the BPASS binary evolution library that at any point in their evolution pass through an evolutionary state similar to the current state of HR 6819.}
    \label{fig:bpass_models}
\end{figure}

We search the $Z=0.02$ BPASS grid and set the following constraints:
\begin{itemize}
    \item $15 < T_{\rm eff,1}/{\rm kK} < 17$
    \item $2.5 < \log g_1 < 3.25$
    %\item $15 < T_{\rm eff,2}/{\rm kK} < 20$
    \item $5 < M_2/M_{\odot} < 10 $
    \item $20 < P/{\rm days} < 60$
    \item $2.7 < \log(L_1/L_{\odot}) < 3.3$
\end{itemize}
We refrained from placing constraints on the temperature or luminosity of the secondary, because the evolutionary models in the BPASS do not follow the secondary's evolution with detailed calculations following mass transfer. Because BPASS uses a grid of discrete models, we allow a relatively large range of periods. 
This search yielded 10 models, with initial primary masses ranging from 4.5 to 7 $M_{\odot}$, initial mass ratios ranging from 0.4 to 0.8, and initial periods of 1.5 to 6.5 days. These are shown in Figure~\ref{fig:bpass_models}.

We then used MESA to calculate binary evolution tracks for models in this region of parameter space, and to investigate how the systems' evolution changes when their parameters are varied. Most of the single-star free parameters are set following the \texttt{inlist\_7M\_prems\_to\_AGB} inlist in the MESA test suite. We use the wind mass loss prescription from \citet{Reimers_1975}. The initial metallicity is set to 0.02. We use a mixing length parameter $\alpha_{\rm MLT} = 1.73$ and an exponential overshoot scheme with overshoot efficiency parameter $f_{\rm ov}=0.014$. The overshooting scheme and motivation for this choice of $f_{\rm ov}$ are described in \citet{Herwig_2000}.

The \texttt{MESAbinary} module is described in detail in \citet{Paxton_2015}. Both stars are evolved simultaneously, with tides and mass transfer taken into account. Roche lobe radii are computed using the fit of \citet{Eggleton_1983}. Mass transfer rates in Roche lobe overflowing systems are determined following the prescription of \citet{Kolb_1990}. Tidal torques are modeled following \citet{Hut_1981}. The orbital separation evolves such that the total angular momentum is conserved when mass is lost through winds or transferred to a companion, as described in \citet{Paxton_2015}. We experimented with a range of mass transfer efficiencies. When mass transfer is not conservative, the mass that escapes the binary is modeled as being lost from the vicinity of the accretor through a fast wind, as described by \citet[][their ``$\beta$'' parameter]{Tauris_2006}.

\begin{figure*}
    \includegraphics[width=\textwidth]{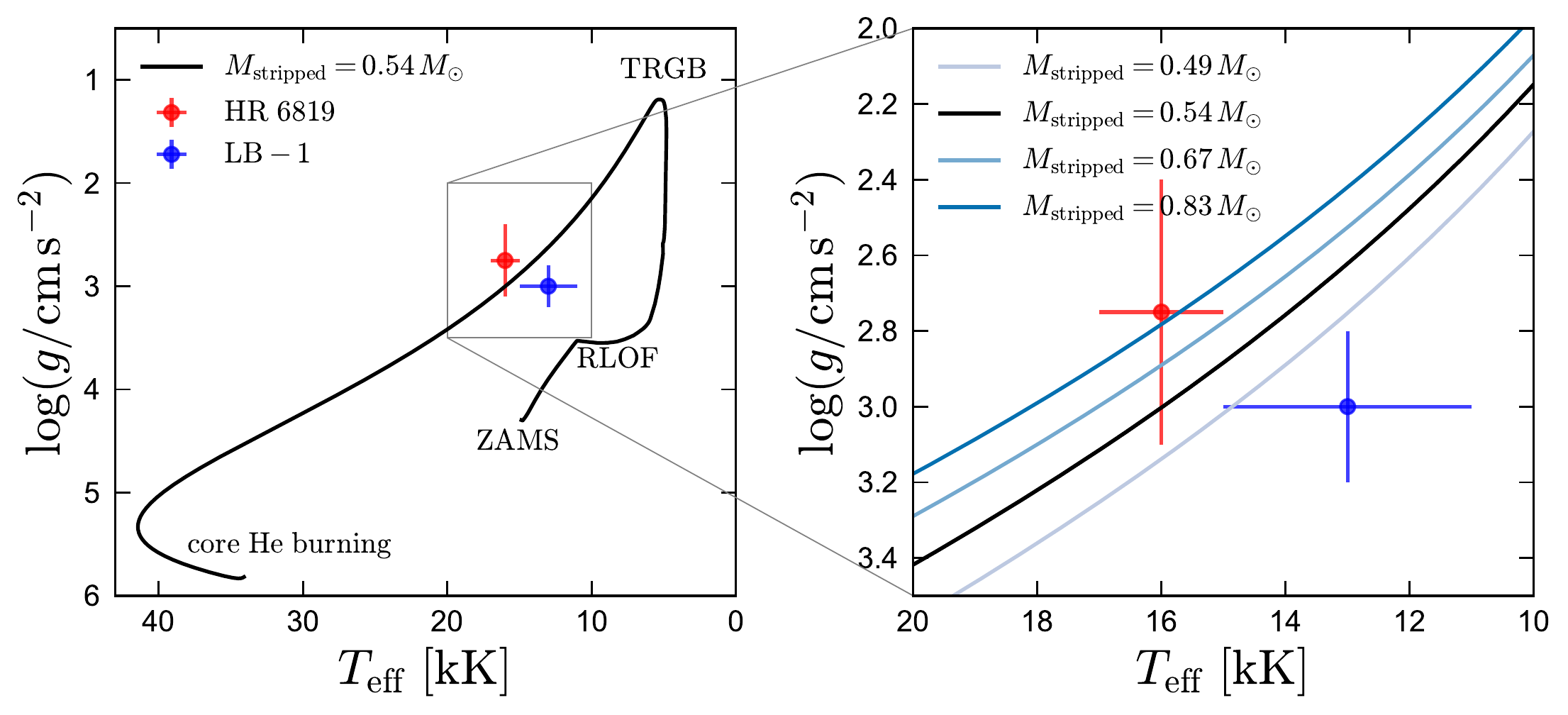}
    \caption{Evolutionary history of the primary for stripped star models similar to HR 6819 and LB-1. Left panel shows a model with initial masses $M_1=4M_{\odot}$ and  $M_2=3M_{\odot}$, and initial period 2.3 days. This produces a stripped star with mass $0.54 M_{\odot}$. We indicate the locations of the zero-age main sequence (ZAMS), first Roche lobe overflow (RLOF), tip of the red giant branch (TRGB), and stable core helium burning. Right panel compares models with a range of stripped star masses, corresponding to initial masses between 3.5 and 7 $M_{\odot}$. }
    \label{fig:hrd_stripped}
\end{figure*}

\begin{table*}
\centering
\caption{Summary of our MESA binary models.}
\begin{tabular}{lllllllllllll}
\hline\hline
%\multicolumn{3}{l}{\bf{MESA models}}  \\ 
$M_{1, \rm init}$ & $M_{2, \rm init}$ & $P_{\rm init}$ & $f_{\rm mt}$ &  $P_{\rm stripped}$ & $M_{\rm stripped}$ &  $M_{\rm He\,core}$ & $M_{\rm env}$ &  $M_{\rm Be}$ & $t_{\rm RLOF}$ & $t_{\rm TRGB}$ & $t_{\rm sdOB}$  & bloated stripped phase lifetime  \\
$[M_{\odot}]$ & $[M_{\odot}]$ & [days] &  &  [days] & $[M_{\odot}]$ & $[M_{\odot}]$ &  $[M_{\odot}]$ &  $[M_{\odot}]$ & [Myr] & [Myr] & [Myr]  &  [$10^5$\,yr]  \\

\hline
3.5 & 2.8 & 2.3 & 1   & 90  & 0.49 & 0.43 & 0.06 & 5.8 & 232 & 269.9 & 272.7 & 2.07  \\
4.0 & 3.0 & 2.3 & 1   & 91  & 0.54 & 0.47 & 0.07 & 6.4 & 162 & 203.8 & 205.5 & 1.29  \\
5.5 & 4.2 & 2.3 & 0.3 & 152 & 0.67 & 0.11 & 0.56 & 5.6 & 71  & 97.1  & 98.7  & 2.20  \\
7.0 & 4.0 & 2.3 & 0.2 & 79  & 0.83 & 0.15 & 0.68 & 5.2 & 41  & 55.8  & 57.2  & 2.27  \\

\hline
\end{tabular}
\begin{flushleft}
$M_{1, \rm init}$, $M_{2, \rm init}$, and $P_{\rm init}$ are the initial masses and period at the zero-age main sequence. $f_{\rm mt}$ is the efficiency of mass transfer, defined such that a fraction $f_{\rm mt}$ of mass lost by the donor is gained by the accretor. $P_{\rm stripped}$ is the orbital period during the bloated stripped star phase, when $2 < \log g < 3.5$ (roughly when the star passes through the inset in Figure~\ref{fig:hrd_stripped}). $M_{\rm stripped}$, $M_{\rm He\,core}$, and $M_{\rm env}$, are the total, core, and envelope mass of the initial primary (the B star) during this period, and $M_{\rm Be}$ is the mass of the companion during this period. $t_{\rm RLOF}$, $t_{\rm TRGB}$, and $t_{\rm sdOB}$ are the system age when the primary first overflows its Roche lobe,  when it reaches the tip of the red giant branch, and when its luminosity is first dominate by He burning.  The ``bloated stripped phase lifetime'' is the time it takes the primary to contract from $\log g =2$ to $\log g = 3.5$ as it moves from the TRGB toward the extreme horizontal branch. 
 
\label{tab:mesa}
\end{flushleft}
\end{table*}

Our MESA models are summarized in Table~\ref{tab:mesa}.
Figure~\ref{fig:hrd_stripped} shows the evolution of 4 models in $T_{\rm eff}-\log g$ space. The left panel shows a model with initial masses $M_1 = 4 M_{\odot}$, $M_2 = 3 M_{\odot}$, and an initial period of 2.3 days. The components of the binary evolve normally until the end of the primary's main sequence lifetime. Shortly before the primary crosses the Hertzsprung gap, it overflows its Roche lobe and begins stable mass transfer to the companion. The primary's envelope expands further as more mass is removed, and the star moves up the red giant branch. At the tip of the red giant branch (TRGB), the temperatures required to maintain hydrogen burning can only be maintained if the envelope contracts. At this point, when the primary's luminosity is of order $1000 L_{\odot}$, the primary leaves the RGB and passes through the region of $T_{\rm eff}$--$\log g$ space where HR 6819 and LB-1 are currently observed. The star then continues to contract and heat until it reaches $T_{\rm eff}\approx 30\,\rm kK$ and $\log g \approx 5.5$, where it settles as a core helium burning hot subdwarf (sdOB star). The core helium burning phase would last of order $10^8$ years, but for most of the models we consider, it is truncated prematurely when the secondary (the Be star) leaves the main sequence and overflows it Roche lobe. 

The right panel compares stripped star models with masses between 0.49 and 0.83 $M_{\odot}$. These are produced from calculations with initial primary masses of 3.5, 4.5, 5.5, and 7 $M_{\odot}$. We reduce the efficiency of mass transfer (which is poorly constrained a priori) as the primary mass increases, so that the mass of the secondary during the stripped star phase is always about 6 $M_{\odot}$. As the total mass of the core increases, its radius and luminosity increase, and $\log g$ decreases. However, the sensitivity of these parameters to mass is relatively weak in this region of parameter space, so HR 6819's atmospheric parameters are consistent with models with a broad range of core masses (and hence initial masses).

\begin{figure}
    \includegraphics[width=\columnwidth]{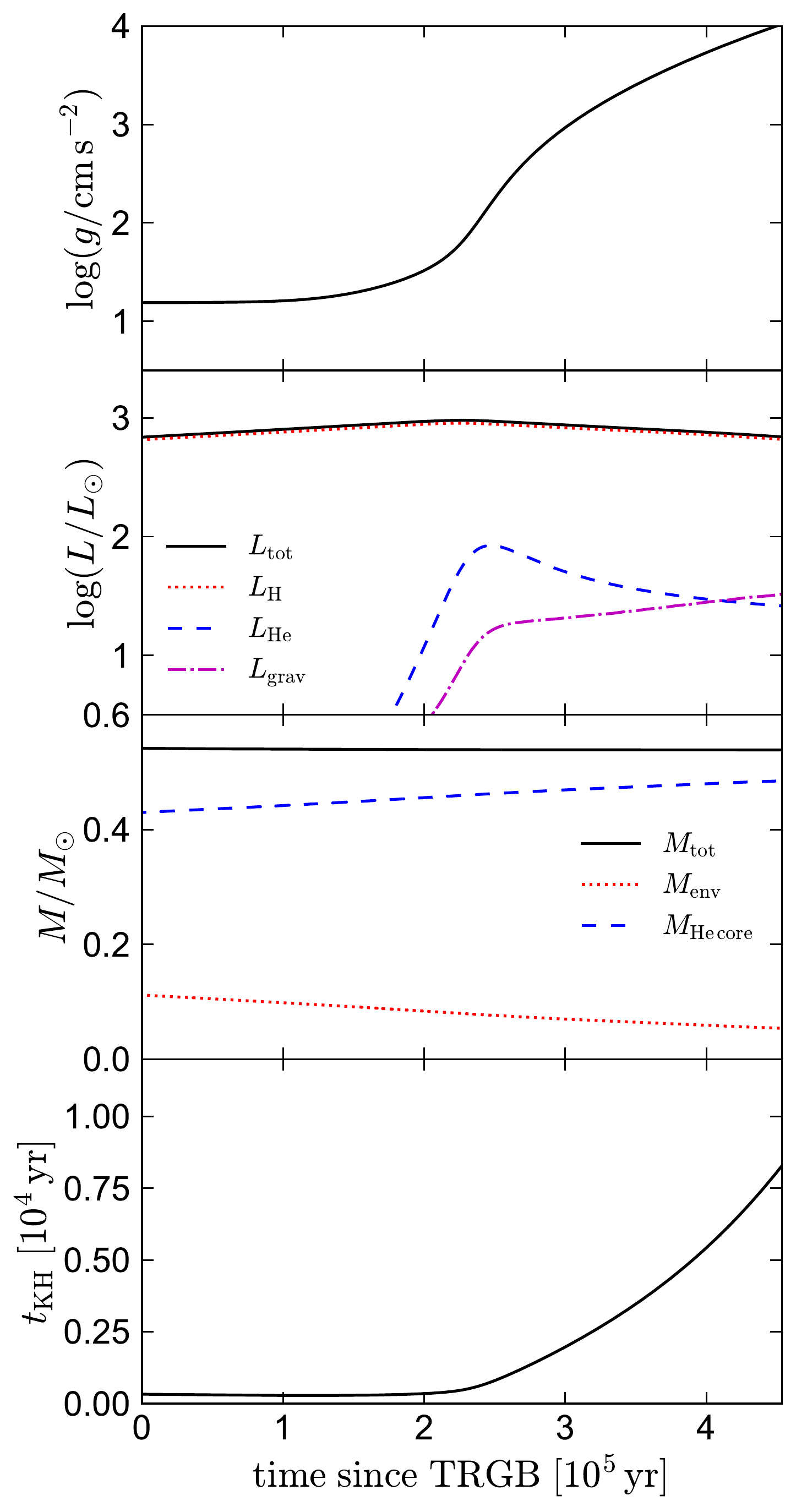}
    \caption{Time-evolution of the MESA model with $M_{\rm stripped}=0.54 M_{\odot}$ during the stripped star phase (black line in Figure~\ref{fig:hrd_stripped}). From top to bottom, panels show surface gravity, luminosity, mass, and Kelvin-Helmholtz timescale. Time is measured since the tip of the red giant branch (i.e. when $\log g$ reaches its minimum). The luminosity is dominated by H shell fusion, which burns through most of the remaining envelope as the star contracts. He fusion is ignited as contraction begins, but does not dominate over shell fusion until $1.5\times 10^6$ years after the TRGB (beyond the time range of the plot). The lifetime of the contraction phase is a few $\times 10^5\,$ years. This is significantly longer than the Kelvin-Helmholtz timescale (bottom panel), because gravitational contraction contributes only a small fraction of the luminosity. }
    \label{fig:time_evol}
\end{figure}

Figure~\ref{fig:time_evol} shows the evolution of the model with  $M_1 = 4 M_{\odot}$ and $M_2 = 3 M_{\odot}$ after the primary begins to contract. We designate the ``bloated stripped star'' phase of the binary's evolution as the phase after the tip of the red giant branch when the primary (the B star) has $\log g$ between 2 and 3.5. This is roughly the period when the primary has the temperature of a B star and is more inflated than a main sequence star of the same temperature. For the 4 models shown in Figure~\ref{fig:hrd_stripped}, the duration of this phase ranges from 1.3 to 2.3 $\times 10^5$ years. The secondary (i.e., the Be star) does not evolve on the timescale shown in Figure~\ref{fig:time_evol}: for all our models, the mass transfer timescale near the TRGB is significantly longer than the thermal timescale of the secondary, so the secondary remains in thermal equilibrium. 

We note that, contrary to what is sometimes assumed in the literature \citep[e.g.][]{Eldridge_2020}, the contraction from the red giant branch to the extreme horizontal branch does not occur on a Kelvin-Helmholtz timescale of the donor star -- the bottom panel of Figure~\ref{fig:time_evol} shows that it is longer by more than a factor of 10. The reason for this is that gravitational contraction is not the dominant source of energy during this phase; hydrogen shell burning and core helium burning both contribute luminosity to slow the star's contraction.

The final period produced by these calculations is somewhat longer ($P=79-152$ days) than the observed $P=40.3$ days in HR 6819. This can be understood as a result of the orbit's expansion during mass transfer. When mass transfer is conservative, the final period can be calculated analytically to be
\begin{align}
    \label{eq:p_final}
    P_{f}=P_{i}\left(\frac{M_{{\rm donor,}i}}{M_{{\rm donor},f}}\times\frac{M_{{\rm accretor},i}}{M_{{\rm accretor,f}}}\right)^{3}.
\end{align}
Here $P_i$ and $P_f$ represent the initial and final period, $M_{{\rm donor,}i}$ and $M_{{\rm donor,}f}$ the initial and final masses of the donor, and $M_{{\rm accretor},i}$ and $M_{{\rm accretor},f}$ the initial and final masses of the accretor. In our calculations $M_{{\rm donor,}i}/M_{{\rm donor,}f}= 7-8.5$, while $M_{{\rm accretor,}i}/M_{{\rm accretor,}f}= 1-2.1$. This causes the final period to be a factor of $\gtrsim 40$ larger than the initial period. The final period according to Equation~\ref{eq:p_final} could be shorter if the initial mass ratio were more unequal ($M_1 \gg M_2)$, but in this case mass transfer eventually becomes dynamically unstable, leading to an episode of common envelope evolution (CEE) and typically inspiral to a much shorter period \citep[e.g.][]{Heber_2016}. 

% see Soberman_1997
No analytic expression for the final period exists for non-conservative mass transfer. How the angular momentum of a binary changes following mass loss depends on how the mass is lost (see \citealt{Soberman_1997}, and \citealt{Tauris_2006}, their Equation 16.18). In \texttt{MESAbinary}, it is possible to specify what fraction of transferred mass is lost from the vicinity of the donor, from the vicinity of the accretor, and from a circumbinary toroid. Particularly near the end of the stripping process, when $M_{\rm donor}/M_{\rm accretor} \ll 1$, the orbital response to mass loss, which is parameterized through the \citet{Tauris_2006} subgrid model, depends quite sensitively on the assumed mass loss fractions for each channel. For our calculations with non-conservative mass transfer, the final period decreases by up to a factor of 10 if we assume mass is lost from a circumbinary toroid rather than from a wind in the vicinty of the accretor. Because we have limited physical intuition for what the efficiency of mass transfer is in this problem or how mass that leaves the binary escapes, the observed period cannot reliably constrain the binary's initial period.

% but Table~\ref{tab:mesa} shows that the final periods for our calculations with mass transfer efficiency $f_{\rm mt} < 1$ are not much shorter than those in the conservative calculations.
%A potential resolution to this tension (i.e., a way to achieve shorter final periods) is provided by torques from the circumbinary disk formed during RLOF \citep[e.g.][]{Artymowicz_1994}, which are thought to drive sdOB binaries formed through stable mass transfer to shorter periods \citep{Vos_2015}.

We note that while luminosity and radius increase monotonically with the total mass of the stripped star in our calculations (right panel of Figure~\ref{fig:hrd_stripped}), this is not necessarily expected to generically be the case. The radius and luminosity of the stripped star depend on the relative masses of the envelope and the He core, with lower envelope masses corresponding to larger radii and luminosities at fixed temperature. For example, \citet{Driebe_1998} calculated evolutionary tracks for low-mass stripped pre-He WDs formed from a $1 M_{\odot}$ progenitor by artificially enhanced mass loss on the RGB. Their $0.41\,M_{\odot}$ track is nearly coincident with our $0.83 M_{\odot}$ track, while their $0.33\,M_{\odot}$ track falls below our $0.49 M_{\odot}$ track and is consistent with the LB-1 atmospheric parameters \citep[see also][]{Irrgang_2020}. In HR 6819, both the dynamical mass and atmospheric parameters of the stripped star are consistent with a scenario in which it is a pre-He white dwarf, with a degenerate core and too low a mass to ever ignite helium. However, our binary evolution calculations do not produce stripped stars with such low mass. Because the stripped star must have initially been the more massive component of the binary, and the total initial mass must have been at least $\gtrsim 6 M_{\odot}$, the initial mass of the stripped star must have been at least $3 M_{\odot}$. Stars with initial masses near $3M_{\odot}$ {\it can} produce stripped He stars with masses of order $0.4 M_{\odot}$, but these ignite He burning on the RGB. Because their cores are not degenerate during the bloated stripped phase, they have thicker hydrogen envelopes, higher $\log g$, and lower luminosity than the \citet{Driebe_1998} models. 

We have not performed an exhaustive search of parameter space with detailed binary evolution calculations, so it is possible that another region of parameter space could produce systems similar to HR 6819 and LB-1, even though no other satisfactory models were found in the BPASS library. Below, we briefly summarize how the MESA models change when the initial conditions or model parameters are perturbed. 
\begin{itemize}
    \item {\it Longer periods}: For longer initial periods, mass transfer begins later, when the primary has expanded more. For $P_{\rm init}\lesssim 20$\,days, evolution is similar to our fiducial models, but stripping occurs over a shorter timescale, the final period is longer, and a somewhat larger fraction on the envelope remains when contraction begins, shifting models downward (toward larger $\log g$) by up to $\approx 0.2\,\rm dex$ in Figure~\ref{fig:hrd_stripped}.
    
    For $20 \lesssim P_{\rm init}/{\rm days} < 150$, mass transfer begins when the envelope is convective and the core mass is too small to support stable mass transfer, leading to dynamical instability and presumably an episode of CEE.  MESA does not include robust treatment of CEE, but it is expected to result in shorter periods than the observed $P=40.3$ days \citep[e.g.][]{Ivanova_2013}.
    
    For $150 \lesssim P_{\rm init}/{\rm days} < 2000$, the primary begins core He burning before RLOF. Stripping then begins on the asymptotic giant branch (AGB), where the larger core mass allows mass transfer to remain dynamically stable \citep[e.g.][]{Chen_2008}. Eventually, the primary is reduced to a carbon-oxygen core with a thin helium envelope, whose contraction is similar to that of a single proto-white dwarf. These models pass through the same region of $T_{\rm eff}-\log g$ space highlighted in Figure~\ref{fig:hrd_stripped}, typically shifted upward (toward smaller $\log g$) somewhat from our fiducial models of the same total mass. However, they contract much more rapidly than our fiducial models, with lifetimes of $\lesssim 10^3$ years. The final periods are also much longer, $P \gtrsim 3000$ days, though we note that these depend sensitively on the assumptions made about how mass is lost from the binary.
    \item { \it Shorter periods }: For shorter periods, RLOF occurs earlier, when the primary is still on the main sequence. For $P_{\rm init} \lesssim 2$ days, the primary loses about half its mass during the initial phase of RLOF, but then essentially continues its evolution as a lower-mass main sequence star. Meanwhile, the now-more-massive secondary evolves faster than the primary, reaching the end of the main sequence before the primary. Such models never resemble HR 6819. 
    \item {\it More unequal mass ratios}: If the initial masses of the two components are too unequal ($M_2/M_1 \lesssim q_{\rm crit}$, where $q_{\rm crit}$ depends on the structure of the donor; e.g. \citealt{Hjellming_1987, Han_2002, Ge_2010}), mass transfer eventually becomes dynamically unstable, leading to an episode of CEE. When mass transfer begins, the envelope is radiative, and mass transfer is stable for a wide range of mass ratios. However, the envelope becomes convective soon after mass transfer begins, and mass transfer then becomes unstable. We note that the secondary is generally not expected to accrete much mass during CEE; this also speaks against an initially low secondary mass.
    
    \item { \it Higher primary masses }: As the mass of the primary increases, the efficiency of mass transfer must approach 0 to avoid making the Be star too massive. If this is achievable physically, then a higher primary mass produces leads to a more massive stripped He star. For our default parameters, $M_{1,\rm init}=10\,M_{\odot}$ leads to a 1.3 $M_{\odot}$ stripped He star. During the contraction phase, the $1.3 M_{\odot}$ model continues the monotonic trend seen in Figure~\ref{fig:hrd_stripped} of lower $\log g$ with increasing mass at fixed $T_{\rm eff}$. 

    \item {\it Lower primary masses}: For primary masses $M_{1,\rm init}\lesssim 3 M_{\odot}$, the total initial mass of the binary is less than its current total mass, so the observed mass of the Be star cannot be reproduced. 
    
    \item {\it More or less overshooting}: The treatment of overshooting changes the mapping between initial mass and stripped star mass. The larger the overshooting efficiency parameter, the larger the convective core on the main sequence, and the larger the He core. Increasing $f_{\rm ov}$ thus leads to higher $M_{\rm stripped}$ for fixed $M_{1,\rm init}$, shifting models toward lower $\log g$ at fixed $T_{\rm eff}$. For the models shown in Figure~\ref{fig:hrd_stripped}, increasing $f_{\rm ov}$ from 0.014 to 0.028 increases $M_{\rm stripped}$ by $\approx 0.12M_{\odot}$ on average, but does not significantly change the lifetime of the ``bloated stripped star'' phase.
    
\end{itemize}

We note that while the models shown in Figure~\ref{fig:hrd_stripped} allow for a consistent solution between the dynamically-inferred mass of the stripped star in HR 6819 and its position in $T_{\rm eff}-\log g$ space, we have not found a consistent solution for LB-1. The $T_{\rm eff}$ and $\log g$ (taken from \citealt{Shenar_2020}) imply a stripped star mass $M_{\rm stripped} \lesssim 0.5 M_{\odot}$, which can be relaxed to $M_{\rm stripped} \lesssim 0.6 M_{\odot}$ if we allow for longer initial periods. However, \citet{Shenar_2020} report a dynamical mass of $(1.5\pm 0.5) M_{\odot}$, which is significantly higher than this. A $1.5 M_{\odot}$ He star is generically difficult to produce in the context of our models (or the BPASS models); if stripping occurs near the Hertzsprung gap, it requires a primary mass of order $12 M_{\odot}$ for our adopted overshooting parameter. Unless mass transfer is highly inefficient, the Be star would end up too massive in such a scenario. Without binary interactions, a single star of solar metallicity must have initial mass $\gtrsim 9 M_{\odot}$ to reach a He core mass of $1.5 M_{\odot}$ prior to core He burning. 

The dynamical mass constraint derived by \citet{Shenar_2020} comes primarily from the H$\alpha$ line. Since spectral disentangling relies on the assumption that the individual spectra of both components are time-invariant, and this assumption likely does not hold in detail for the H$\alpha$ line of LB-1 due to phase-dependent disk irradiation by the B star \citep[][]{Liu_2019}, it is possible that the dynamical mass measurement of the B star is biased and the true mass is lower.
If the mass measurement is reliable, it would imply a significantly higher envelope mass ($M_{\rm env}/M_{\rm He\,core} \sim 1$) than produced by our models, since the luminosity during the stripped phase is set mainly by the He core mass. Three of the five confirmed Be + sdOB binaries have estimated masses above $1.0 M_{\odot}$ \citep{Wang_2017}, so a higher stripped-star mass is not beyond the realm of possibility. This is, however, at least partially a selection effect, since higher-mass sdOBs are hotter and easier to detect \citep[e.g.][]{deMink_2014, Wang_2017, Schootemeijer_2018}. We also note that the secondaries in these systems have higher masses than in HR 6819 ($(9-12)\,M_{\odot}$), allowing for higher initial masses for the primaries. 

\section{Summary and Discussion}
\label{sec:discussion}
We have shown that HR 6819 is a binary containing a rapidly rotating Be star and an undermassive B star with mass of order $0.5\,M_{\odot}$. Radial velocity measurements from double-lined spectra establish that the two luminous components of the binary are orbiting one another (Figure~\ref{fig:velocities}). There is thus no need to invoke an unseen 3rd object to explain the orbital motion of the B star, as proposed by \citet{Rivinius_2020}. We propose that the B star is the core of a star with initial mass of order $5\,M_{\odot}$ whose envelope was recently stripped by its companion near the tip of the red giant branch. In this scenario, the B star is currently in a short-lived (few $\times 10^5$ years) phase of contraction toward the extreme horizontal branch, where it will become a core helium burning sdOB  star \citep[e.g.][]{Heber_2016}. In this scenario, HR 6819 is a progenitor system for Be + sdOB binaries, of which at least six are known \citep[e.g.][]{Gies_1998, Koubsky_2012, Peters_2013, Peters_2016, Wang_2017, Dulaney_2017, Chojnowski2018} and another dozen candidates have been identified \citep{Wang_2018}.
Recent accretion of the B star's envelope would then explain the near-critical rotation of the Be star.

HR 6819 is quite similar to the binary LB-1 \citep{Liu_2019}, which has recently been the subject of much discussion. In \citet{ElBadry_2020}, we speculated that the emission lines in LB-1 might come from a circumbinary disk around a BH and a stellar companion. But spectral disentangling by \citet{Shenar_2020} has since revealed the presence of a Be star in the spectrum, similar to the case in HR 6819. The B star in LB-1 would then also be a recently stripped He star, with no BH in the system.  

Further modeling will be required to pin down the mass and current evolutionary state of these systems. Our MESA calculations produce stripped stars with masses, temperatures, and surface gravities comparable to HR 6819 and LB-1, but there is not yet full agreement between the dynamically- and evolutionarily-inferred masses for LB-1. For both systems, {\it Gaia} DR3 astrometry (including binary astrometric model fits) and ground-based interferometry are expected to prove useful in further constraining the system parameters.   

\subsection{Implications for the formation of Be stars}
\label{sec:implications}
We now consider what the existence of the HR 6819 system implies for the broader Be star population. 

HR 6819 is bright enough ($G=5.22$, $V=5.36$) to be seen with the naked eye. Estimating the rate of Be + stripped star binaries based on HR 6819 is challenging both because the search that led to its discovery does not have a well-defined selection function and because Poisson errors are large when only one object is known. We nonetheless make a crude estimate of the frequency of HR 6819-like binaries below.

Roughly 15\% of all field B stars are Be stars \citep{Jaschek_1983, Zorec_1997}. There are about 1500 B stars brighter than $G=6$, corresponding to about 225 Be stars.\footnote{We estimate the number of B stars brighter than $G=6$ based on {\it Gaia} photometry. There are 1275 sources in {\it Gaia} DR2 with $-0.3 < G_{\rm BP}-G_{\rm RP} < 0.1$, a color range corresponding to $T_{\rm eff}=(11-30)\,\rm kK$ without extinction. We inflate this to 1500 to account for extinction (which is modest for most naked-eye stars) and incompleteness in {\it Gaia} DR2 for bright stars. We obtain a similar estimate based on the Hipparcos catalog. The Be star catalog from \citet{Jaschek_1982} contains 141 sources brighter than $V=6$ but does not claim to be complete.} Most stars this bright have been studied in some detail, so it seems a reasonable assumption that HR 6819 is the brightest object in its class, and that there are at most a few stars brighter than $G=6$ in its current evolutionary state. 

Once the stripped star contracts, HR 6819 will likely continue to appear as a Be star for a few $\times\,10^7$ years -- a few hundred times the lifetime of the stripped star phase -- depending on the mass of the Be star. This implies that for every system like HR 6819, there should be a few hundred binaries consisting of a Be star and a faint sdOB or WD companion. Such binaries certainly exist, but they are challenging to detect because the sdOB typically contributes at tiny fraction of the light in the optical. 5 of the 6 confirmed Be + sdOB binaries are brighter than $G=6$ and were discovered via UV spectra. Recently, \citet{Wang_2018} identified another dozen candidate Be + sdOB systems using IUE spectra, all of which are brighter than $G=7$. \citet{Schootemeijer_2018} estimated that there should be 30-50 yet undiscovered Be + sdOB binaries for each confirmed system of a given magnitude, which corresponds to 150-250 Be + sdOB binaries brighter than $G=6$. \citet{Raguzova_2001} estimated through population synthesis simulations that about 70\% of Be stars formed through binary evolution should have faint white dwarf companions, while 20\% should have sdOB companions. No white dwarf companions have yet been unambiguously detected.

Given our estimate that there are only a few hundred Be stars brighter than $G=6$, the existence of HR 6819 is consistent with a scenario in which a majority of Be stars form through this evolutionary channel. This is an appealing possibility, because the origin of the Be phenomenon is still not fully understood. It is certainly tied to rapid rotation, but there is considerable debate about how Be stars were spun up in the first place \citep[e.g.][]{Rivinius_2013}. Accretion of mass from a companion is one possibility. There is, however, considerable uncertainty in the frequency of HR 6819-like binaries. Given $N=1$ objects discovered among 225 Be stars, the 1-sigma Poisson confidence interval for the fraction of Be stars in this evolutionary phase is $(0.08-1.5)$\%. If we take the mean lifetime of a Be star to be $5\times 10^7$ years and the mean lifetime of the stripped star to be $2\times 10^5$ years, the lower limit of 0.08\% implies that at least 20\% of Be stars went through a phase like HR 6819.

HR 6819 and LB-1 are currently the only known objects in their class. Given that LB-1 is more than 6 magnitudes fainter than HR 6819 and was discovered in an multi-epoch RV survey of only 3000 objects (without selection on color or spectral type), there are probably significantly more similar systems with magnitudes between HR 6819 and LB-1 that have yet to be discovered. We estimate that there are of order $10^4$ Be stars brighter than $G = 12$.\footnote{This estimate is also based on {\it Gaia} DR2. Because reddening is more severe for the $G< 12$ sample, we select candidate B stars as objects with  $-0.3 < G_{\rm BP}-G_{\rm RP} < 0.3$ and $M_{\rm G} < 1.5$; this yields $9.5\times 10^4$ B star candidates, of which we again assume 15\% are Be stars.} Given the rate estimate from HR 6819, this implies between 8 and 150 Be + bloated stripped star binaries brighter than $G = 12$, a large fraction of which should be discoverable to wide-field spectroscopic surveys such as SDSS-V. While both HR 6819 and LB-1 were only identified as candidate stripped stars after significant spectroscopic study, additional candidates could likely be identified from a single medium-resolution spectrum as objects exhibiting both the emission lines of a classical Be star and narrow absorption lines from the companion (indicative of low $v \sin i$). Normal Be stars viewed pole-on would be a contaminant to such a search, but these can usually be identified from the shape of their emission lines \citep{Hummel_2000, Rivinius_2013}, and they are rare due to geometric effects.

With LB-1 and HR 6819 demoted, the list of detached stellar mass BH candidates remains short. The most secure detached BH candidates to date are likely those discovered in globular clusters with MUSE \citep{Giesers_2018, Giesers_2019}, but the evolutionary history of these objects is likely quite different from that of BH binaries in the field. No luminous companion has been discovered to the giant--BH binary candidate reported by \citet{Thompson_2019}, but its status as a BH is not uncontroversial \citep{vandenHeuvel_2020, Thompson_2020}. We note that the other two  BH candidates on the field also contain Be stars \citep{Casares_2014, Khokhlov_2018}. These systems are not likely to contain low-mass stripped stars, because in them it is the Be star that is clearly RV variable. Be stars do not constitute a dominant fraction of all massive stars, and there is no strong bias toward Be stars in the selection function of multi-epoch RV surveys. If these objects do contain BHs, then their status as Be stars is likely to be a result of accretion from the BH progenitor. 

In the final stages of this manuscript's preparation, a similar analysis of HR 6819 was completed by \citet{Bodensteiner_2020}. Overall, there is excellent agreement between their results and ours. The only point of disagreement, the helium abundance of the B star, is discussed in Appendix~\ref{sec:He_enrichment}.

\section*{Data availability}
All the spectroscopic data is publicly available through the ESO archive. The TLUSTY/SYNSPEC spectral models are available at \href{http://tlusty.oca.eu/}{http://tlusty.oca.eu/}. MESA is publicly available at \href{http://mesa.sourceforge.net/}{http://mesa.sourceforge.net/}. Other data is available upon reasonable request of the corresponding author.

\section*{Acknowledgements}
We are grateful to the referee, Philip Dufton, for a constructive report.
We thank Thomas Rivinius, Dan Weisz, Todd Thompson, and Andrea Antoni for helpful discussions. KE acknowledges support from an NSF graduate research fellowship and a Hellman fellowship from UC Berkeley. We thank Geoff Tabin and In-Hei Hahn for their hospitality during the preparation of this manuscript. 

%%%%%%%%%%%%%%%%%%%%%%%%%%%%%%%%%%%%%%%%%%%%%%%%%%



\bibliographystyle{mnras}

%\begin{thebibliography}{99}
%\end{thebibliography}

%%%%%%%%%%%%%%%%%%%%%%%%%%%%%%%%%%%%%%%%%%%%%%%%%%

%%%%%%%%%%%%%%%%% APPENDICES %%%%%%%%%%%%%%%%%%%%%

\appendix

\section{Stellar parameters anchored to evolutionary tracks and distance}
\label{sec:distance_anchor}

\begin{figure}
    \centering
    \includegraphics[width=\columnwidth]{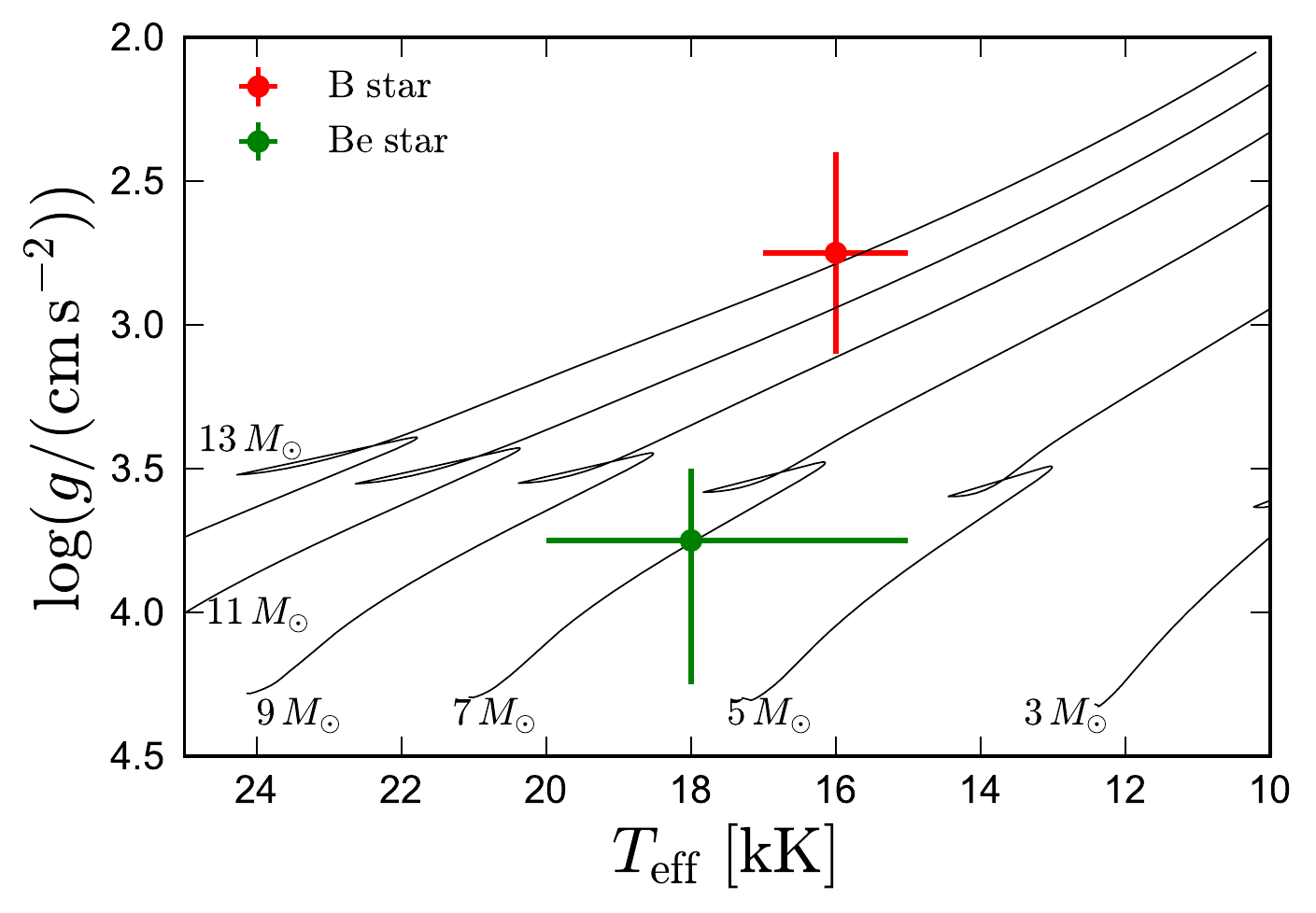}
    \caption{Atmospheric parameters for the two components of HR 6819 compared to MIST evolutionary tracks for normal (i.e. non-stripped) stars. If the B star were a normal star, its $T_{\rm eff}$ and $\log g$ would imply a mass of at least $9\,M_{\odot}$, roughly 20 times the mass we infer dynamically. For the Be star, the atmospheric parameters imply a mass between 5 and 8 $M_{\odot}$.}
    \label{fig:kiel_diag}
\end{figure}

Figure~\ref{fig:kiel_diag} compares the atmospheric parameters of both components of HR 6819 to MIST evolutionary tracks. As discussed in Section~\ref{sec:lum}, the temperature and surface gravity of the B star would imply a mass of at least $9\,M_{\odot}$ if it were a normal star, but this is inconsistent with the measured system luminosity (Figure~\ref{fig:cmd}). The atmospheric parameters of the Be star imply a mass in the range of 5 and 8\,$M_{\odot}$. For a single star, this would correspond to an age range of 30 to 80 Myr, but we caution that mass transfer between the two components is likely to make these age estimates unreliable. 

To estimate the mass and radius of the Be star, we construct a grid of evolutionary tracks with a mass spacing of 0.05 $M_{\odot}$. We sample each track with 500 points spaced uniformly in time. We thus make no attempt to account for the different IMF probability or lifetime of each track (which would downweight higher-mass tracks), or the observational selection function (which would downweight lower-mass tracks). We then retain samples with probability proportional to the likelihood of their $T_{\rm eff}$ and $\log g$ given the measured constraints on the Be star's atmospheric parameters, approximating the likelihood function as a two-sided Gaussian. This effectively selects points from the evolutionary tracks that are close to the Be star in $T_{\rm eff}-\log g$ space. We use the resulting samples to estimate the mass, radius, and luminosity of the Be star, and to propagate forward the uncertainty in other parameters in Table~\ref{tab:system} that depend on the Be star's mass.  

The system parameters listed in Table~\ref{tab:system} are anchored to the mass of the Be star estimated from its $T_{\rm eff}$ and $\log g$. The constraints can be tightened somewhat if the distance, $d=(310\pm 60)\,\rm pc$, is taken into account. To this end, we add a term to the likelihood function that compares the predicted $V-$band apparent magnitude of the evolutionary tracks to the measured $V=5.36$, assuming a $V$-band flux ratio of $f_{{\rm Be}}/f_{{\rm tot}}=0.52\pm0.05$ and extinction $A_V=0.42$. The resulting constraints are listed in Table~\ref{tab:distance_and_tracks}. They are consistent with the constraints in Table~\ref{tab:system}, but the uncertainties are somewhat smaller. At fixed $T_{\rm eff}$, a constraint on luminosity translates directly to a constraint on radius. When combined with the measured $\log g$, this translates to an improved constraint on mass.

\subsection{System parameters independent of evolutionary tracks}
An estimate of the stellar parameters of both components that is independent of evolutionary tracks can be calculated from the apparent magnitude, distance, and flux ratio. The $V$-band apparent magnitude of HR 6819 is 5.36. Given extinction $A_{V}=0.42$ mag, total bolometric correction $\rm BC = -1.5\pm 0.15$,\footnote{For solar-metallicity MIST models, $T_{\rm eff}$ values of 16\,kK and 18\,kK correspond to bolometric corrections of $-1.35$ and $-1.64$, with weak dependence on $\log g$.} and distance $d= 369\pm 34$\,pc, the absolute bolometric magnitude of HR 6819 is $M_{\rm bol} = -4.39^{+0.26}_{-0.24}$ mag. This corresponds to a bolometric luminosity 
\begin{align}
    \label{eq:Lbol}
L_{{\rm bol,tot}}&=10^{0.4\left(M_{{\rm bol}, \odot}-M_{{\rm bol}}\right)}L_{\odot}\\
    \label{eq:Lbol2}
    &=4500_{-1000}^{+1100}L_{\odot},
\end{align}
where $M_{{\rm bol}, \odot}=4.74$.

Given the effective temperatures of the two components, their respective uncertainties, and $(47\pm 5)$\% flux contribution from the B star at $\lambda = 4000\,\textup{\AA}$, the fraction of the total light contributed by the B star is $f_{\rm B,bol}=0.43_{-0.07}^{+0.08}$. The individual luminosities are then:
\begin{align}
\label{eq:LbolB}
\log(L_{{\rm bol,B}}/L_{\odot})&=3.28^{+0.13}_{-0.13}\\
\label{eq:LbolBe}
\log(L_{{\rm bol,Be}}/L_{\odot})&=3.40^{+0.12}_{-0.12}.
\end{align}

The radius of each component can then be calculated as $R=\sqrt{L_{{\rm bol}}/\left(4\pi\sigma_{\rm SB} T_{{\rm eff}}^{4}\right)}$, where $\sigma_{\rm SB}$ is the Stefan Boltzmann constant. This yields 
\begin{align}
\label{eq:Rb}
R_{{\rm B}}&=5.7_{-0.9}^{+1.1}\,R_{\odot}\\
\label{eq:Rbe}
R_{{\rm Be}}&=5.3_{-1.1}^{+1.9}\,R_{\odot}.
\end{align}

Finally, the mass of each component can be calculated as $M=gR^{2}/G$, yielding 
\begin{align}
\label{eq:Mb}
M_{{\rm B}}&=0.65_{-0.38}^{+0.93}\,M_{\odot}\\
\label{eq:Mbe}
M_{{\rm Be}}&=6.9_{-4.0}^{+15.7}\,M_{\odot}.
\end{align}

Equations~\ref{eq:LbolB}-\ref{eq:Mbe} are fully consistent with the constraints in Tables~\ref{tab:system} and~\ref{tab:distance_and_tracks}, which are anchored on the fit of the temperature and gravity of the Be star with evolutionary tracks. Equations~\ref{eq:LbolB}-\ref{eq:Mbe} are also consistent with, but independent of, the dynamically measured mass ratio.  We adopt the parameters anchored to evolutionary tracks as our fiducial values due to their tighter constraint on the component masses. %We note, however, that the constraints on the component radii in Equations~\ref{eq:Rb} and \ref{eq:Rbe} are actually somewhat stronger than those derived from evolutionary tracks. 

%\begin{table}
%\centering
%\caption{System parameters anchored to distance only}
%\begin{tabular}{lll}
%\hline\hline
%
%
%Be star mass &  $M_{\rm Be}\,[M_{\odot}]$ & $4.8^{+11.6}_{-3.0}$ \\ 
%Be star radius & $R_{\rm Be}\,[R_{\odot}]$ & $4.5^{+1.8}_{-1.2}$  \\ 
%Be star bolometric luminosity & $\log(L_{\rm Be}/L_{\odot})$ & %$3.25^{+0.18}_{-0.21}$ \\ 
%B star mass &  $M_{\rm B}\,[M_{\odot}]$ & $0.46^{+0.71}_{-0.28}$ \\ 
%B star radius & $R_{\rm B}\,[R_{\odot}]$ & $4.8^{+1.3}_{-1.1}$  \\ 
%B star bolometric luminosity & $\log(L_{\rm B}/L_{\odot})$ & %$3.13^{+0.18}_{-0.22}$ \\ 
%
%\hline
%\end{tabular}
%\begin{flushleft}
%
%\label{tab:distance_and_tracks}
%\end{flushleft}
%\end{table}
%
%
%
\begin{table}
\centering
\caption{System parameters anchored to both distance and evolutionary tracks. Uncertainties are 1$\sigma$ (middle 68\%).}
\begin{tabular}{lll}
\hline\hline

%\multicolumn{3}{l}{\bf{Be star parameters}}   \\ 
Be star mass &  $M_{\rm Be}\,[M_{\odot}]$ & $7.0^{+1.0}_{-0.8}$ \\ 
Be star radius & $R_{\rm Be}\,[R_{\odot}]$ & $5.3^{+0.9}_{-0.7}$  \\ 
Be star bolometric luminosity & $\log(L_{\rm Be}/L_{\odot})$ & $3.45^{+0.14}_{-0.16}$ \\ 
Be star fraction of critical rotation & $v_{\rm rot}/v_{\rm crit}$ & $0.83^{+0.13}_{-0.11}$ \\ 

Orbital inclination & $i\,[\rm deg]$ & $31.7^{+1.7}_{-1.5}$  \\
Semi-major axis  & $a$ [R$_{\odot}$]   &  $96\pm 4$ \\
B star mass &  $M_{\rm B}\,[M_{\odot}]$ & $0.49^{+0.24}_{-0.22}$ \\ 
Be star outer disk radius & $R_{\rm outer}\,[R_{\odot}]$ & $33^{+4}_{-3} $ \\
Be star inner disk radius & $R_{\rm inner}\,[R_{\odot}]$ & $7.9^{+1.9}_{-1.8} $ \\

\hline
\end{tabular}
\begin{flushleft}

\label{tab:distance_and_tracks}
\end{flushleft}
\end{table}

\section{Spectral disentangling and determination of \texorpdfstring{$K_{\rm Be}$}{} }
\label{sec:disentangle_details}

Spectral disentangling solves for the two single-star spectra that can best reproduce all the observed composite spectra {\it for an assumed set of relative RVs at each epoch}. If the RVs are not known a priori, they can be solved for in tandem with the single-star spectra by searching for the RVs that allow the disentangled single-star spectra to best reproduce all the observed composite spectra. In the case of HR 6819, the problem is simplified by the fact that the narrow lines of the B star allow a preliminary estimate of its RVs to be made directly from the composite spectra.

We optimize for the RVs in two steps. In the first step, we assume that the velocity of the B star at each epoch is the velocity predicted by the orbital solution of \citet{Rivinius_2020}. We then step through a grid of RV semi-amplitudes for the Be star, $K_{\rm Be}$, calculating the optimal disentangled spectra for each value. The choice of $K_{\rm Be}=0\,\rm km\,s^{-1}$ corresponds to a stationary Be star, as would be expected in a triple scenario.\footnote{Note that there is no need to try different center-of-mass velocities: because we do not use any template spectra, the absolute RV scale is arbitrary. } We choose the optimal $K_{\rm Be}$ and corresponding spectral decomposition as the value that minimizes the total $\chi^2$ summed over all observed spectra. 

With the disentangled spectra from the first step in hand, we use them as templates to fit the composite spectra for the RVs at each epoch, as described in Section~\ref{sec:velocites}. This revealed intrinsic scatter of $s\approx 4\,\rm km\,s^{-1}$ for the RVs of the B star (see  Section~\ref{sec:velocites}).

In the second step, we repeat the disentangling, again stepping through a grid of $K_{\rm Be}$ values. Rather than setting the RVs of the B star to the values predicted by the \citet{Rivinius_2020} orbital solution, we now set them to the values measured in the previous step. The best-fit values of $K_{\rm Be}$ for a given spectral region do not change much between the first and second step, but the quality of the fit improves somewhat in the second step.

We perform spectral disentangling and determine the optimal $K_{\rm Be}$ for 10 different regions of the spectrum, each centered on an emission or absorption line with clear contributions from both components.  Treating different lines independently allows us to test whether different ions trace different kinematic components, and to quantify the uncertainty in the $K_{\rm Be}$ values we infer.

\begin{figure}
    \includegraphics[width=\columnwidth]{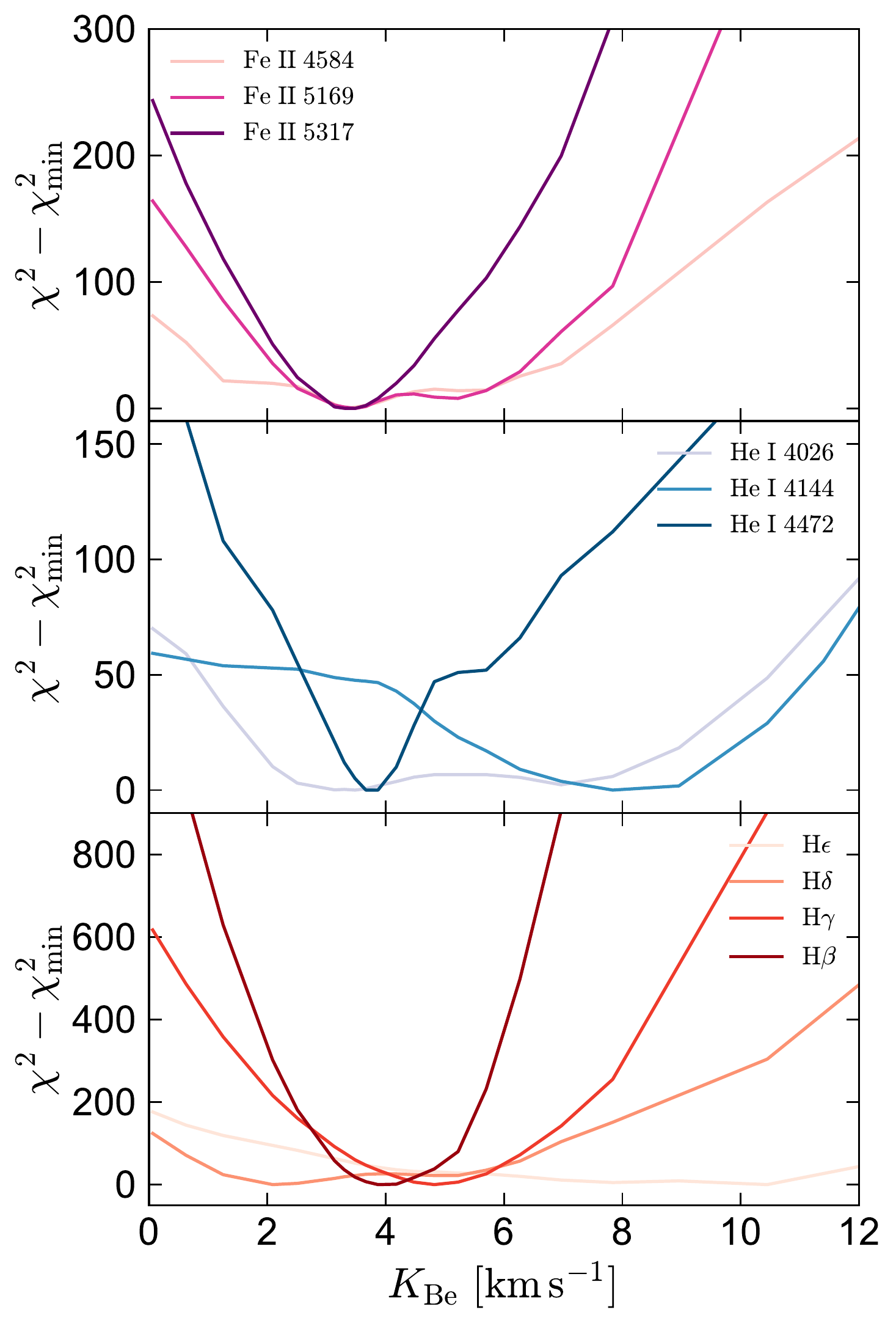}
    \caption{Total $\chi^2$ versus RV semi-amplitude of the Be star assumed during spectral disentangling. Top, middle, and bottom panels respectively show Fe II lines, He I lines, and H Balmer lines.   }
    \label{fig:chi2_kbe}
\end{figure}

Figure~\ref{fig:chi2_kbe} shows the total $\chi^2$ as a function of $K_{\rm Be}$ for the 10 lines. For easy comparison, we subtract the minimum $\chi^2$ for each line.  The Fe II lines in the top panel are dominated by emission in the Be star spectrum (e.g. Figure~\ref{fig:emission_lines}). The He I lines (middle panel) are dominated by absorption. The Balmer lines\footnote{We do not fit the H$\alpha$ line, because the shape of the emission varies between epochs. That is, there is no two-component spectral decomposition for which a good fit of all the observed spectra can be obtained. Such variability is common for Be stars \citep[e.g.][]{Rivinius_2013}; it may or may not be related to the companion.} contain both absorption and emission, with a larger emission contribution toward redder wavelengths (Figure~\ref{fig:disentangled_spectra}). 

The 10 spectral regions we consider have optimal $K_{\rm Be}$ ranging from 2 to 10 $\rm km\,s^{-1}$. Of these, 7 have optimal values between 3 and 6 $\rm km\,s^{-1}$. None of the lines have minimal $\chi^2$ values associated with $K_{\rm Be}=0\,\rm km\,s^{-1}$, as would be expected for a stationary Be star.  The scatter in $K_{\rm Be}$ values inferred from different lines, including lines tracing the same ion, is larger than would be expected if the formal uncertainties in $K_{\rm Be}$, as quantified by $\chi^2(K_{\rm Be})$ for a single line, were reliable. This is most likely a consequence of small errors in continuum normalization as well as time-variability in the flux ratio, which violate the underlying assumptions of the spectral disentangling algorithm \citep[see][]{Hensberge_2008}.

\begin{figure}
    \includegraphics[width=\columnwidth]{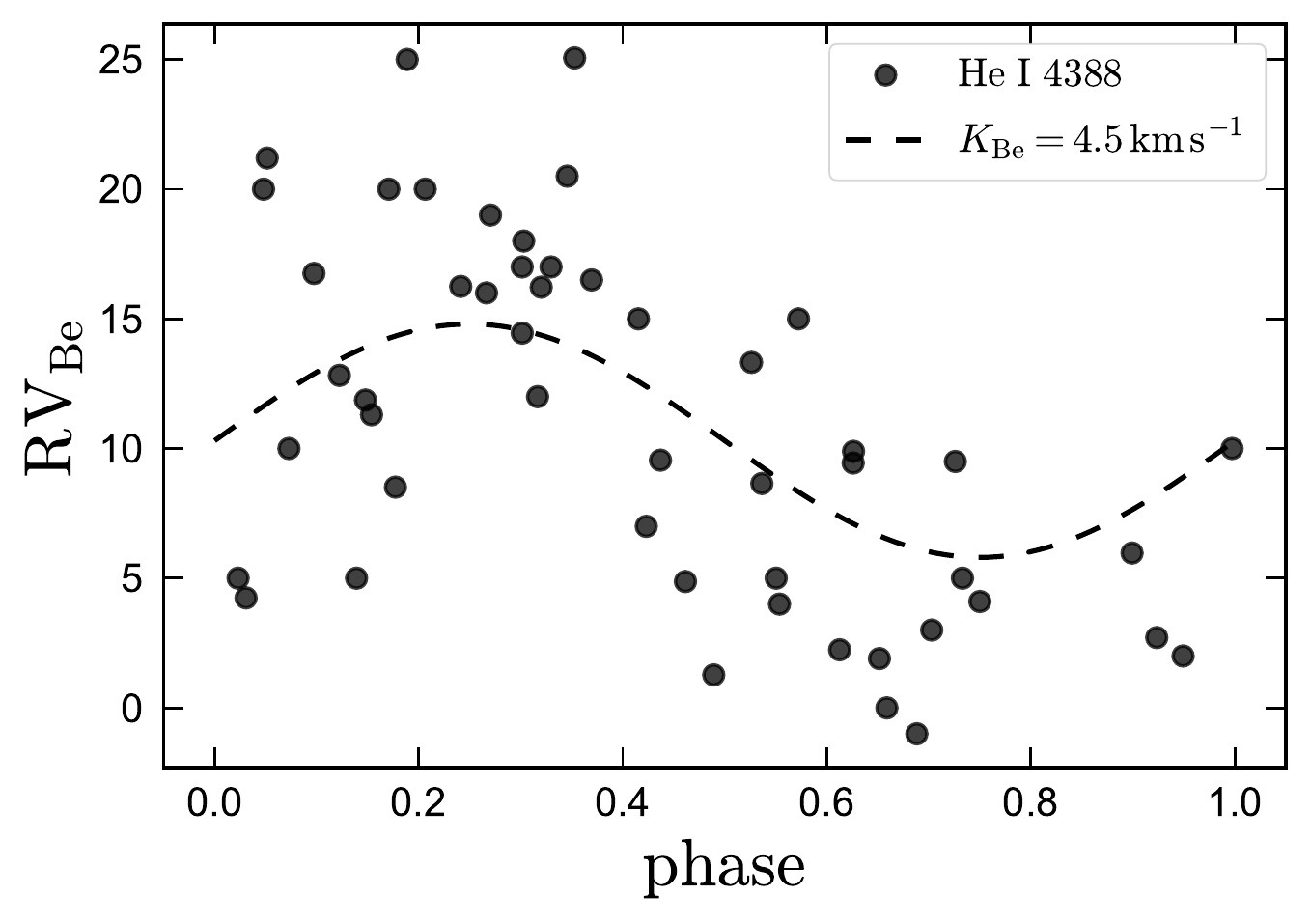}
    \caption{RVs for the Be star measured by fitting the He I absorption line at 4388\,\textup{\AA} with TLUSTY/SYNSPEC model spectra. Dashed line shows the prediction for a circular orbit with $K_{\rm Be} = 4.5\,\rm km\,s^{-1}$. The typical RV uncertainty ($\sigma_{\rm RV,Be}\approx 5\,\rm km\,s^{-1}$) is substantial compared to the amplitude of the RV variability, but there is clear phase-dependent variability.}
    \label{fig:synthetic_rvs}
\end{figure}

We do not find systematic difference between lines from different ions, or emission vs. absorption lines. This suggests that the emission lines, which are presumed to originate in a disk around the Be star, move coherently with the star. Considering the 10 spectral regions shown in Figure~\ref{fig:chi2_kbe}, as well as the RVs obtained when fitting a synthetic spectral template (Figure~\ref{fig:synthetic_rvs}), we adopt $K_{\rm Be}=(4.5\pm 2)\,\rm km\,s^{-1}$ as our fiducial value. We then fix $K_{\rm Be}=4.5\,\rm km\,s^{-1}$ in disentangling the full spectrum. 

The shape of the disentangled spectra depends very weakly on the value of $K_{\rm Be}$. Choosing $K_{\rm Be}=0\,\rm km\,s^{-1}$ makes both the emission and absorption lines of the disentangled Be star spectrum slightly broader (by of order 0.1\,\textup{\AA}) than in our fiducial solution. Broader lines suppress the RV variability of the Be star, but this leads to a worse overall fit to the composite spectra. The inferred $T_{\rm eff}$ and $\log g$ of the two components are not sensitive to the adopted value of $K_{\rm Be}$, at least over the range $0 \leq K_{\rm Be}/(\rm km\,s^{-1}) < 10$.

\begin{figure}
    \includegraphics[width=\columnwidth]{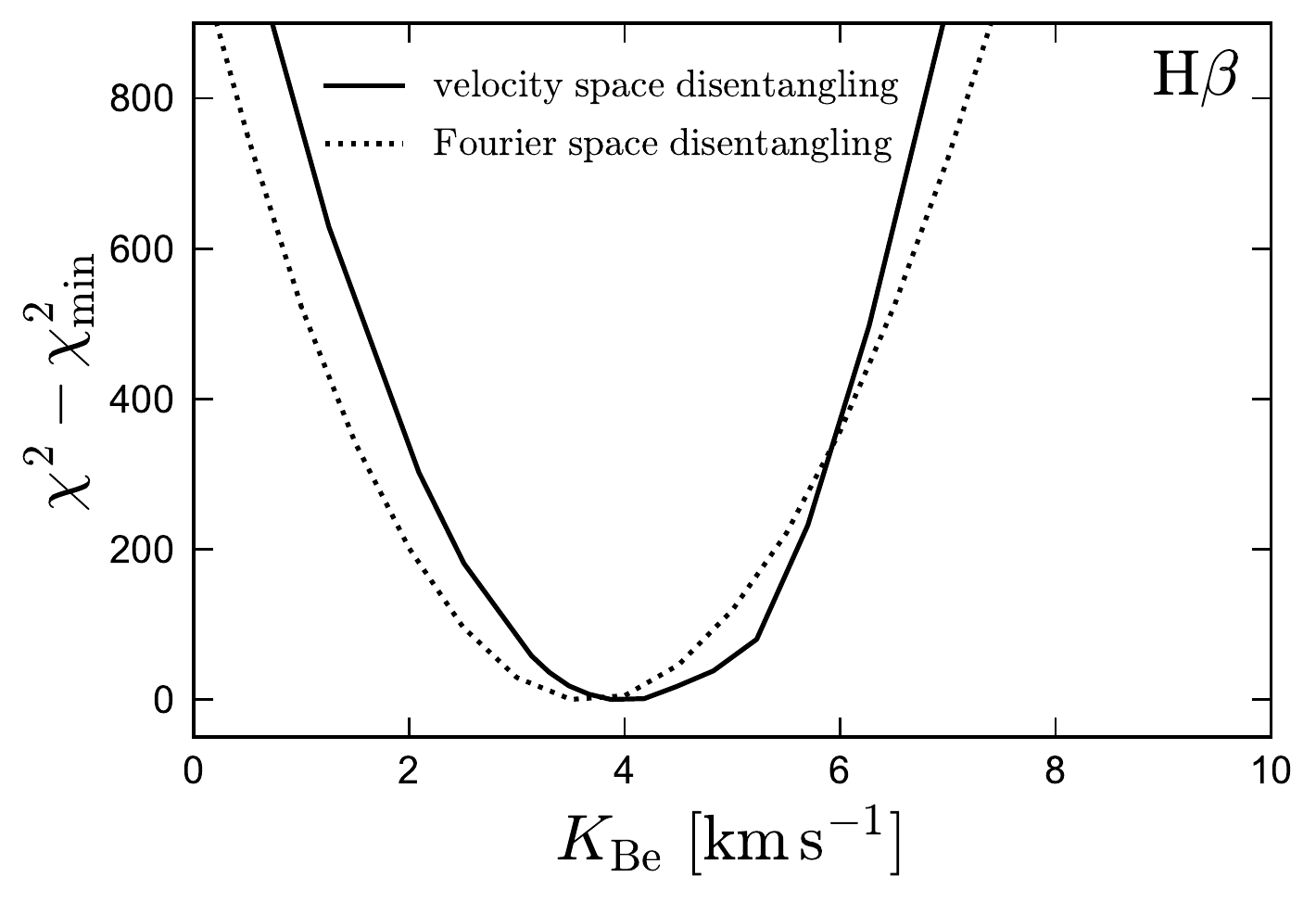}
    \caption{Constraints on $K_{\rm Be}$ from the H$\beta$ line. We compare results obtained using the velocity-space spectral disentangling code \texttt{CRES} (solid line) and the Fourier-space code \texttt{fd3} (dotted line).} 
    \label{fig:cres_vs_fd3}
\end{figure}

In addition to our primary spectral decomposition using \texttt{CRES}, we also experimented with using the Fourier-space spectral disentangling code \texttt{fd3} \citep[also referred to as \texttt{FDBinary} in the literature;][]{Ilijic_2004b, Ilijic_2017}, which uses the algorithm described by \citet{Hadrava_1995} to separate composite spectra. The disentangled spectra produced by the two codes are very similar. Figure~\ref{fig:cres_vs_fd3} compares the two codes' constraints on $K_{\rm Be}$ from the $\rm H\beta$ line. Under idealized circumstances, the velocity-space and Fourier-space spectra disentangling algorithms are mathematically identical \citep[e.g.][]{Hensberge_2008}. The most significant practical difference in the implementations for our purposes is that \texttt{fd3} requires the RVs of both components to follow Keplerian orbits, while \texttt{CRES} allows them to be specified manually at each epoch. The \texttt{fd3} solution thus does not model the scatter in the B star's RVs, which is accounted for in the \texttt{CRES} solution. 

We also experimented with fitting RVs for both components using synthetic spectral templates. Figure~\ref{fig:synthetic_rvs} shows the best-fit RVs of the Be star obtained by jointly fitting the RVs of both components at each epoch, using the best-fit TLUSTY/SYNSPEC spectra for both components and fitting the He I absorption line at 4387.9 \textup{\AA}. Because the synthetic spectra do not perfectly reproduce the observed spectra, the narrow lines in the Be star's spectrum are washed out by rotation, and many of the Be star's lines are significantly contaminated by emission, we find that we obtain more stable RVs when fitting a narrow wavelength range than when fitting the full spectrum simultaneously. The RVs measured this way show larger scatter than those measured from the disentangled spectra (Figure~\ref{fig:velocities}), but they still show clear periodic modulation, the amplitude of which is consistent with the amplitudes we measure from spectral disentangling. The RVs we measure for the B star are in good agreement with those measured in Section~\ref{sec:velocites}.

\section{B star variability}
\label{sec:variability}
As discussed in Section~\ref{sec:velocites}, the measured RVs of the B star do not perfectly follow the predictions of a Keplerian orbit, but exhibit additional scatter with amplitude $\sigma_{\rm RV}\approx 4\,\rm km\,s^{-1}$. Here we show that this variability is accompanied by variability in the B star's line profiles. 

\begin{figure}
    \includegraphics[width=\columnwidth]{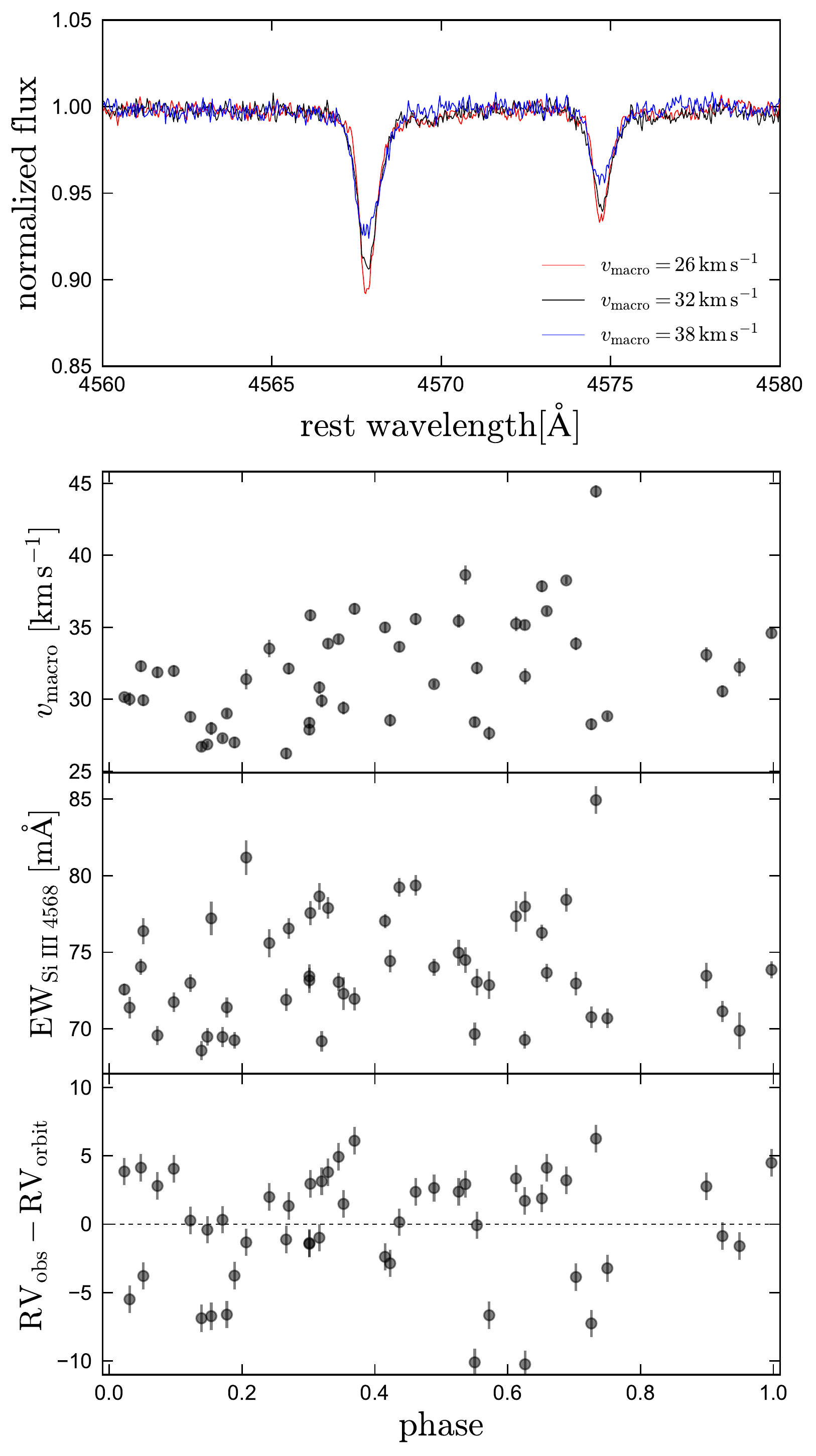}
    \caption{Top panel shows three single-epoch spectra of HR 6189, highlighting two Si III lines for which the absorption is dominated by the B star. The depth and width of the lines varies appreciably between epochs. Legend indicates the best-fit Gaussian $v_{\rm macro}$, which quantifies the width of the lines. Bottom three panels show $v_{\rm macro}$, equivalent width, and RV deviation from a Keplerian orbit, all measured for the Si III 4568 line, in all 51 epochs. These quantities vary between epochs, but do not vary systematically with orbital phase. }
    \label{fig:line_diagnostics}
\end{figure}

The top panel of Figure~\ref{fig:line_diagnostics} shows the line profiles of two Si III lines from three single-epoch spectra of HR 6819, which are chosen to showcase the range of profiles found in the dataset. Weak and narrow metal lines such as these are ideal for characterizing the B star's variability because contamination from the Be star -- besides dilution by its continuum -- is negligible. All three epochs are shifted to rest frame. The depth and width of the lines vary visibly between epochs. We quantify the line width with $v_{\rm macro}=\sqrt{2}\sigma$, where $\sigma$ is the dispersion of the best-fit Gaussian broadening profile (see Appendix~\ref{sec:rot_vs_vmac}). 

The three lower panels of Figure~\ref{fig:line_diagnostics} show $v_{\rm macro}$, equivalent width, and RV deviation from a Keplerian orbit, measured from the Si III 4568 line across all 51 epochs. The epoch-to-epoch variation in $v_{\rm macro}$ is of order 10\%, with a maximum-to-minimum range of almost a factor of 2. The typical variation in equivalent width is somewhat smaller, of order 5\%. The RV scatter is $4\,\rm km\,s^{-1}$, consistent with our finding in Section~\ref{sec:velocites}. There is no clear trend with orbital phase. Like the photometric variability (see \citealt{Rivinius_2020}), the line profile variability exhibits power on timescales from $\approx 0.4$ days to several days, and is not dominated by a single period.

The observed line profile variations cannot be understood as a result of photometric variability of the Be star, which would change the flux ratio -- and thus, line depth and equivalent width --- but not $v_{\rm macro}$. It thus seems natural to connect the B star's line profile variations with the observed photometric variability of HR 6819. The period and amplitude of the photometric variability are consistent with a scenario in which they are due primarily to the B star (see below). We note, however, that photometric variability is also common among Be stars, so we cannot rule out the possibility that some of the observed photometric variability is due to the Be component.

The period, photometric variability amplitude, RV scatter, temperature, and luminosity of the B star in HR 6819 are all typical of slowly pulsating B-type (SPB) stars \citep[e.g.][]{Waelkens_1991, Waelkens_1998}. Indeed, the prototypical SPB star, 53 Persei, has $T_{\rm eff}=17$\,kK, $\log(L/L_{\odot})=2.9$, and exhibits pulsation-driven RV variations with peak-to-peak amplitude of about $15\,\rm km\,s^{-1}$, and photometric peak-to-peak amplitude of order 0.2 mag, quite similar to the B star in HR 6819 \citep{Smith_1984, Chapellier_1998}. SPB stars undergo non-radial $g$-mode pulsations, which are excited by the $\kappa$ mechanism acting on the metal opactiy bump for $T_{\rm eff}=10-25$\,kK \citep[e.g.][]{Dziembowski_1993}. Line profile variations in SPB stars are primarily due to variations in the velocity field at the stellar photosphere; variations due to changes in $T_{\rm eff}$ are usually negligible \citep{Buta_1979, Townsend_1997, Aerts_2000}.

Photometric and spectral variability violate the assumptions of the spectral disentangling algorithm. When variability is purely photometric, it is straightforward to account for a time-dependent luminosity ratio \citep[e.g.][]{Ilijic_2004, Hensberge_2008}. In the presence of line profile variations, however, there is no unique spectral decomposition, so the disentangled spectra represent a time-average of the component spectra. We carried out tests with mock data to assess the effects of unmodeled line profile variability similar to that found in HR 6819 on spectral disentangling. Because the variability is not systematically phase-dependent, we find that it does not lead to systematic biases in $K_{\rm Be}$ or in the width or depths of the disentangled line profiles, but it does reduce the total goodness-of-fit.

\section{Rotation vs. macroturbulent broadening}
\label{sec:rot_vs_vmac}

\begin{figure}
    \includegraphics[width=\columnwidth]{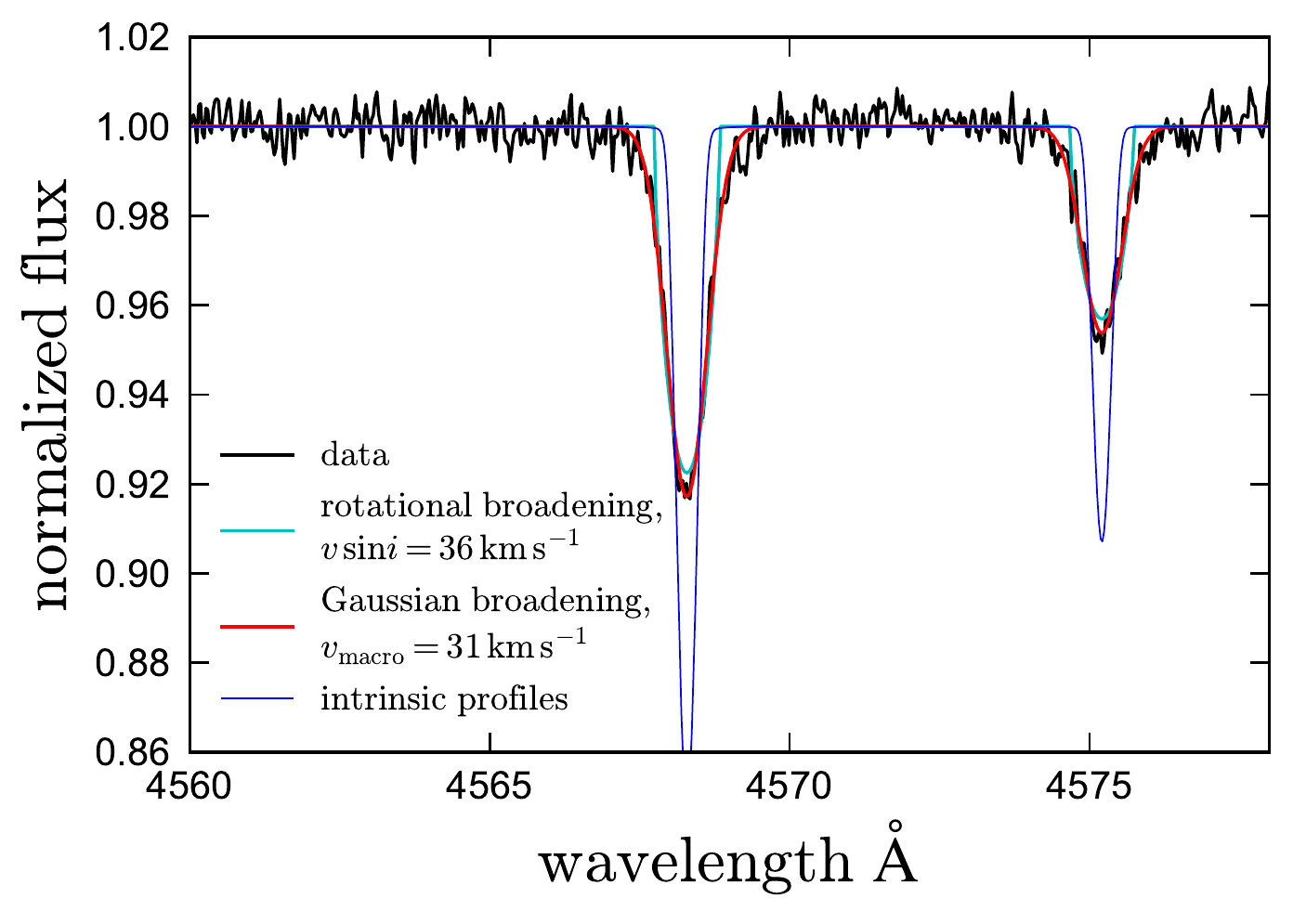}
    \caption{Black line shows an observed spectrum of HR 6819, highlighting two Si III lines for which the absorption is dominated by the B star. Blue line shows a model spectrum with no rotational or macroturbulent broadening and $v_{\rm mic}=10\,\rm km\,s^{-1}$. Cyan and red lines show fits to the line profiles assuming purely rotational and Gaussian broadening, respectively. The observed line profiles have broader wings than predicted for pure rotational broadening and are reasonably well-fit by Gaussian profiles. This indicates that non-rotational broadening contributes significantly to the observed line widths.}
    \label{fig:rot_vmac}
\end{figure}

The metal lines of the B star component of HR 6819 have a typical FWHM of $\rm 50\,\rm km\,s^{-1}$. This is relatively narrow for a B star, but it is larger than the expected thermal width given the B star's spectral type in the absence of rotation or macroturbulent broadening ($\rm FWHM \approx 10\,\rm km\,s^{-1}$). 

Figure~\ref{fig:rot_vmac} shows line profiles of the Si III 4568 and 4575 \textup{\AA} lines in a single-epoch ($\phi \approx 0.55$) spectrum of HR 6819. These lines are ideal for characterizing the B star's line broadening because they are intrinsically narrow and spectral disentangling suggests that the contribution of the Be star to them is negligible. We compare the observed line profiles to the predictions of a TLUSTY/SYNSPEC model ($T_{\rm eff}=16\,\rm kK$ and $\log g=2.75$) with no rotational or macroturbulent broadening, and to line profiles broadened only by macroturbulence or rotation. The rotational broadening profile is taken from \citet{Gray_1992}, using a linear limb darkening coefficient $\epsilon=0.5$. The macroturbulent broadening profile is modeled as a Gaussian with $\sigma = v_{\rm macro}/\sqrt{2}.$ Instrumental broadening has negligible effects on the predicted line profiles. Rotational broadening alone reproduces the observed line profiles poorly. Disagreement between the observed and modeled profiles is most obvious in the line wings, which for the rotational profiles extend only to $\pm v\,\sin i$. The observed line wings are less steep and extend farther from the line center. This is not a result of contamination from the Be star: given its $v \sin i \approx 180\,\rm km\,s^{-1}$, any absorption from it is spread over a significantly wider wavelength range, $\Delta \lambda > 5\,\rm \textup{\AA}$.

Purely Gaussian broadening does reproduce the observed line profiles reasonably well. An isotropic Gaussian broadening profile is not well-motivated physically \citep[e.g.][]{Gray_1992, simon_daiz_2014}, but it is a serviceable simplification. The physical basis of non-rotational broadening in B stars is not well understood. As it is modeled here, macroturbulence represents any broadening mechanisms besides rotation, not necessarily turbulence. Given the observed photometric variability and line profile variations in HR 6819, it is likely that pulsation-driven fluctuations in the surface velocity field of the B star contribute to the broadening. Such fluctuations are not expected to produce strictly Gaussian broadening profiles \citep[e.g.][]{Aerts_2000, Aerts_2014}, but in practice we find that Gaussians provide a reasonably good fit to the line profiles of uncontaminated metal lines in HR 6819 at all epochs, though the best-fit values of $v_{\rm macro}$ change somewhat from epoch to epoch (Figure~\ref{fig:line_diagnostics}). The conclusion that rotation alone produces line profiles with too-steep wings also holds at all epochs.

Because non-rotational broadening likely dominates over rotational broadening, we cannot reliably estimate $v\,\sin i$ of the B star from its line profiles. By fitting the observed Si III 4568 line profiles with a combination of rotational and macroturbulent broadening \citep[e.g.][]{Ryans_2002, simon_daiz_2014}, we can limit the projected rotation velocity to $v\sin i < 20\,\rm km\,s^{-1}$. If the B star is tidally synchronized, the limits on its radius and the binary's inclination (Table~\ref{tab:system}) would imply $v\sin i = 3.1^{+1.8}_{-1.2}\,\rm km\,s^{-1}$. 

\section{Helium enrichment}
\label{sec:He_enrichment}
\begin{figure*}
    \centering
    \includegraphics[width=\textwidth]{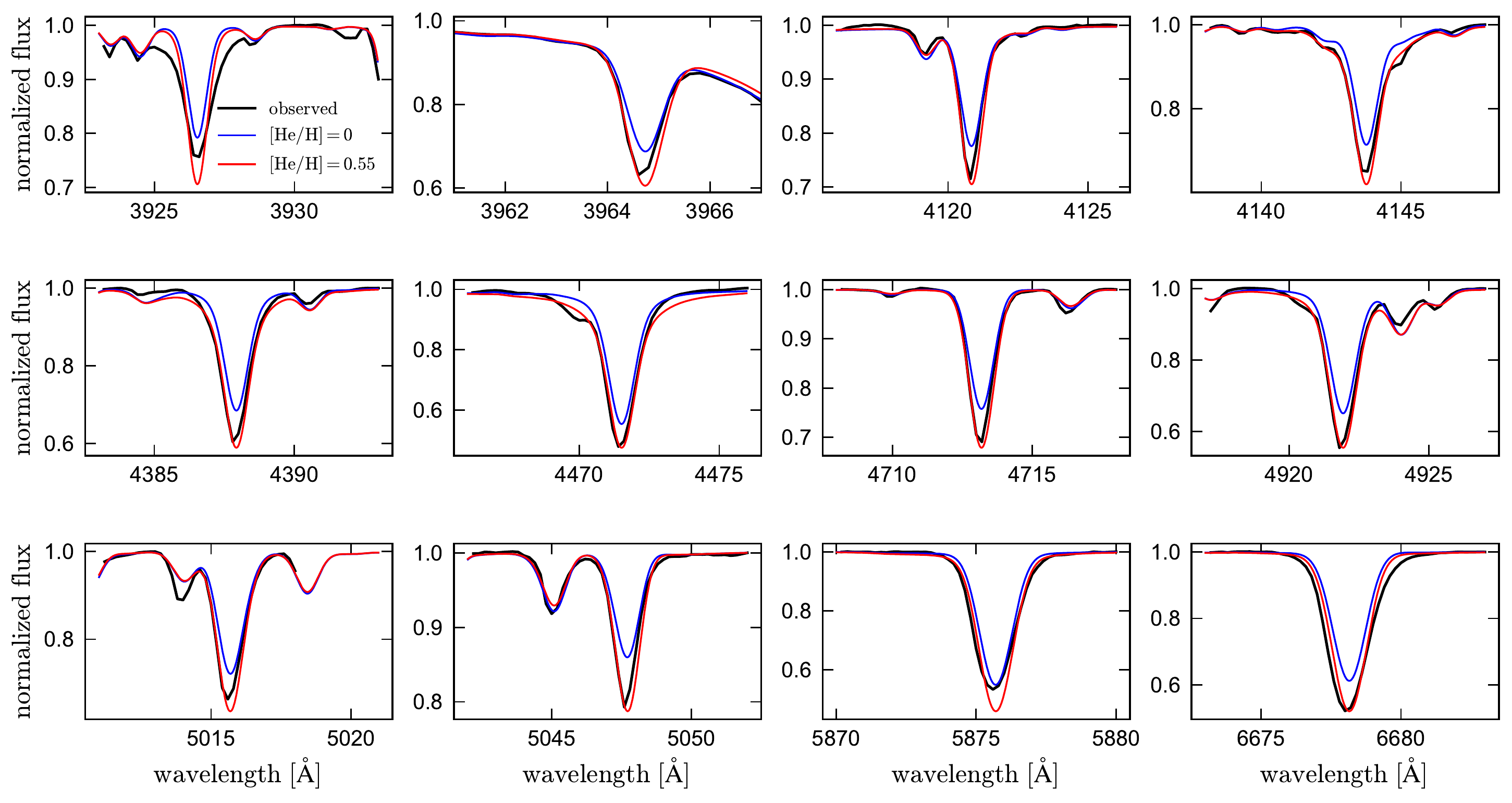}
    \caption{Strong helium lines of the B star. Black line shows the disentangled spectrum of the B star. Blue shows a TLUSTY/SYNSPEC model with $T_{\rm eff} = 16\,\rm kK$, $\log g = 2.75$, $v_{\rm mic}=10\,\rm km\,s^{-1}$, and solar helium abundance, $n_{\rm He}/n_{\rm H}=0.1$. Red line shows a model with the same atmospheric parameters but 0.55 dex higher helium abundance; i.e., $n_{\rm He}/n_{\rm H} = 0.35$. The helium-enhanced model provides a much better fit.}
    \label{fig:he_enrich}
\end{figure*}

As discussed in Section~\ref{sec:abundances}, we find that the surface of the B star is enriched in helium. Evidence for this is shown in Figure~\ref{fig:he_enrich}, which compares the strong He lines in the disentangled spectrum to TLUSTY/SYNSPEC models with solar helium abundance (blue) as well as models with $n_{\rm He}/n_{\rm H} = 0.35$ (red). The latter is 3.5 times the solar value (0.55 dex enhancement) and corresponds to a $\approx 55\%$ helium mass fraction. The helium-enriched models clearly provide a better fit. We note that while a 55\% helium mass fraction is considerably higher than expected for a normal hydrogen-burning star, it is lower than predicted by the MESA models we analyze in Section~\ref{sec:evol}, which predict mass fractions of $\approx$80\%.

\citet{Bodensteiner_2020} also analyzed the spectra of HR 6819. Their inferred atmospheric parameters and  conclusions about the nature of the system are broadly consistent with ours. However, they did not find evidence of helium enrichment. The different conclusions are driven primarily by different choice of atmospheric models: their abundance constraints are based on PoWR models presented in  \citet[][]{Hainich2019}, while we use TLUSTY/SYNSPEC models. Comparing the models directly, we find that at fixed $T_{\rm eff}$, $\log g$, and $v_{\rm mic}$, and He abundance, the PoWR models predict significantly stronger He lines, such that solar-abundance PoWR models indeed provide a good fit to the He lines in HR 6819. We defer a more detailed analysis of the differences between the two atmospheric and spectral models to future work. 

%%%%%%%%%%%%%%%%%%%%%%%%%%%%%%%%%%%%%%%%%%%%%%%%%%

% Don't change these lines
\bsp	% typesetting comment
\label{lastpage}
\end{document}